\documentclass[sigconf,nonacm]{acmart}

\usepackage{graphicx} 
\usepackage{booktabs}
\usepackage{xparse}
\usepackage{array,tabularx,xltabular}
\newcolumntype{L}[1]{>{\raggedright\arraybackslash}p{#1}}
\newcolumntype{Y}{>{\raggedright\arraybackslash}X}
\usepackage{algorithm}
\usepackage{algpseudocode}
\usepackage{subcaption}
\usepackage{tabularx}
\usepackage{tcolorbox}
\tcbuselibrary{listings, skins, breakable} 
\usepackage{siunitx}
\usepackage{cryptocode} 
\usepackage{enumitem}
\usepackage{float}

\usepackage{xcolor} 

\newcommand{\sample}{\xleftarrow{\$}}
\newcommand{\Baigang}[1]{\textcolor{red}{[Baigang: #1]}}

\title{G-Lox: Group-Adaptive, Privacy-Preserving Bridge Distribution with Two-Party Computation}

\author{Baigang Chen}
\affiliation{%
  \institution{University of Minnesota}
  \city{Twin Cities}
  \state{MN}
  \country{USA}}
\email{chen9464@umn.edu}

\author{Nicholas Hopper}
\affiliation{%
  \institution{University of Minnesota}
  \city{Twin Cities}
    \state{MN}
  \country{USA}}
\email{hoppernj@umn.edu}

\begin{document}
\begin{abstract}
Distributing information about unlisted relays ({\em bridges}) to users of an Internet blocking evasion system
is a key unsolved problem for resilient access under strong network information controls.   This is due to the tension
between allowing access to users while preventing blocking authorities from enumerating bridges, and is made more
difficult by the requirement to retain no metadata.  We present \textbf{G-Lox} (\emph{group-adaptive Lox}), a bridge-distribution system that preserves Lox-style distributor blindness while enabling hidden, stateful adaptation at the group level. G-Lox places adaptive assignment logic behind a two-server privacy wall: no single server learns group identifiers or group-to-bridge assignments, while private state access and state-dependent updates are carried out using two-server DPF/FSS protocols and secure two-party computation. This supports blockage reporting, transport-aware reassignment, and privacy-preserving group splitting without exposing sensitive assignment information.

We evaluate G-Lox through both system measurements and policy simulation. Our C++/EMP implementation over real TCP sockets shows that the client-visible overhead of private state access remains small: across state sizes up to \(2^{16}\), communication stays in the low-KiB range per end-to-end iteration. At \(M=1024\), the client sends \(1{,}968\) bytes, receives \(1{,}280\) bytes, and completes an iteration in about \(0.25\)~s. Memory use is also modest, staying near \(5\)~MB on the client and below \(18\)~MB (\(9.79\)~MB at \(M=1024\)) on the state servers in our experiments. At the policy level, simulations with group-specific blocking and Sybil enumeration show that G-Lox consistently improves robustness over Lox- and rBridge-like baselines among systems that maintain broad issuance. Overall, our results show that stronger privacy for bridge-state access can be achieved with practical overhead while improving robustness under adaptive network information control.

\end{abstract}




\maketitle

\section{Introduction.}
Resilience to Internet content blocking (also called network information controls) remains a growing demand, especially in settings where users increasingly face platform blocking, network filtering, and selective interference with communication services. In many regions, network information controls restrict access to major platforms and messaging systems, forcing users to rely on evasion tools, such as VPNs~\cite{alshalan2015survey}, Tor~\cite{dingledine2004tor} or other proxy services, to communicate and access information. However, Internet blocking authorities in some regions can respond to this evasion by observing and blocking publicly-known entry points to these services.  In this case users need alternative, unlisted channels to reach the network. In practice, this role is played by unlisted proxies (called \emph{bridges} in Tor~\cite{dingledine2006design}), often combined with pluggable \emph{transports} that attempt to evade filtering~\cite{fifield2015blocking}. A key challenge in deploying bridges is the inherent tradeoffs in bridge {\em distribution}: a distributor must assign bridges to users while limiting enumeration by blocking authorities, adapting to blocking, and balancing load --- all while leaking as little metadata as possible.

While deployed systems mostly use rate-limiting to combat enumeration and sidestep the other aspects of bridge distribution, a series of research systems~\cite{proximax,wang2013rbridge,douglas2016salmon,nasr2019enemy,tulloch2023lox} have made steady progress toward more effective enumeration defenses based on social-graph type defenses. In these systems, users accumulate "credit" when
their assigned bridges remain available, and users with enough credit can invite additional users.
Systems such as Lox~\cite{tulloch2023lox} and related distributor-side designs, including Salmon~\cite{douglas2016salmon}, show that one can combine anonymous access with cryptographic credentials~\cite{chase2014algebraic} to enforce access policies while keeping the public-facing distributor blind to user identity and linkability signals. At a high level, the distributor in Lox can verify that a request for a bridge is authorized, yet it should not learn which bridge was assigned, nor should it learn stable group identifiers that let it correlate requests across time. 

However, there remain some limitations induced by the gap between these designs' assumptions and the behavior of real network information controls.
First, all previous designs in this model implicitly rely on bridges to detect and report blocking, but in practice, a bridge often cannot reliably know when it is blocked, especially when this status might differ across blocking regions.
Second, network information control is frequently transport-specific and region-specific~\cite{ling2013tor}, so to be most effective, a distributor needs to track per-region per-transport outcomes and rotate transports accordingly.
Third, in systems that assign bridges to groups of related users,
a distributor must handle growth: if a group becomes too large, it should be split to balance load, yet this splitting itself is a stateful operation that can introduce new linkability if handled naively by the distributor. 

In this paper we seek to address these limitations by designing a new bridge-distribution algorithm and protocol implementation, with the following goals: 
\begin{enumerate}
  \item Distributor blindness: As in Lox and rBridge, the public-facing distributor should learn neither which bridge a user is assigned nor which hidden group the user belongs to.

  \item User-driven blockage recovery: Unlike prior designs that implicitly rely on bridges to self-detect blocking, we do not assume bridges can reliably determine when they are blocked. Instead, blockage is signaled by users: a user privately requests a fresh assignment when their current bridge fails by reporting blockage, and the server reassigns a new bridge when reports reach a certain limit. 

  \item Blocking-region specific behavior: users should be assigned
  to (hidden) groups that experience similar blocking behavior.

  \item Transport-aware adaptation: Bridges are associated with a set of transports/types.  The system should track outcomes per group and per transport, and shift a group away from transports that appear degraded for that group. 

  \item Privacy-preserving load balancing: When a hidden group grows beyond a configured threshold, the system should split it into two roughly equal successor groups so that load can be balanced across bridges, without revealing co-membership or group structure to the distributor.
  
\end{enumerate}
To achieve those goals, we adapt the Lox design with additional cryptographic mechanisms to support additional functionality. We present Group-Adaptive Lox (\textbf{G-Lox}), a bridge distribution design that places all group-indexed state and assignment logic behind a two-server privacy wall
\footnote{G-Lox can also stand for "Goldilocks"~\cite{marshall1988goldilocks},  as the user can try different types of transport till it is "just right".}. In our design, two non-colluding state servers jointly compute decisions using 2PC~\cite{goldreich2019play,yao1986generate} on secret-shared inputs and maintain per-group state via private key--value access using two-server distributed point function/function secret sharing (DPF/FSS) point queries~\cite{boyle2015function,boyle2016function}. All assignment outputs are returned exclusively to the user inside an end-to-end encrypted payload, preventing the distributor $D$ from learning bridge selections or stable group identifiers. We summarize our contribution as follows.

\begin{enumerate}
\item \textbf{G-Lox design.}
We present G-Lox, a privacy-preserving bridge-distribution system that retains the core Lox guarantee that the public-facing distributor learns neither users' group membership nor their bridge assignments, while still enabling stateful adaptation. G-Lox places group-indexed state and decision logic behind a two-server privacy wall and private group-state access is realized with two-server DPF/FSS queries. This supports blockage reporting, transport-aware reassignment, and privacy-preserving group splitting without exposing sensitive assignment structure.

\item \textbf{Microbenchmark evaluation.}
We implement the G-Lox backend in C++17 using EMP and evaluate it over real TCP sockets. Our benchmark includes private state queries, 2PC-based adaptation logic, and token-gated two-server DPF-PIR for directory redemption. Across state sizes \(M\in\{2^{10},\ldots,2^{16}\}\), client-visible communication stays in the low-KiB range per end-to-end iteration. Memory use remains modest, with client memory near \(5\)~MB and state-server memory below \(18\)~MB in our experiments.

\item \textbf{Policy-level simulation.}
We develop a policy-level simulator for group-specific blocking, Sybil-driven enumeration, and report-based adaptation, and compare G-Lox against Lox-, rBridge-, and Salmon-like baselines. In stressed settings, G-Lox achieves the strongest robustness among high-issuance systems. In the main zig-zag scenario, it reaches \(74.6\%\) day-30 success with \(100\%\) issuance, compared with \(38.9\%\) for Lox and \(44.7\%\) for rBridge-like baselines.
\end{enumerate}

\section{Related Work.}
\subsection{Background in Bridge Distribution}
\paragraph{Onion routing and Tor.}
Our work builds on the onion-routing family of low-latency anonymity systems, which provide anonymous, socket-style connections by wrapping traffic in layered encryption and relaying it through a chain of nodes, so that each hop sees only its immediate neighbors and local routing context. Classic onion routing formalizes this proxy-based architecture and its resistance goals against eavesdropping and traffic-analysis adversaries~\cite{reed2002anonymous, goldschlag1999onion,camenisch2005formal}. Tor instantiates and extends this approach at Internet scale with circuit construction, directory-based relay discovery, and operational mechanisms that improve deployability (e.g., forward security and exit policies)~\cite{dingledine2004tor, jansen2016safely}.

\paragraph{Bridges and Bridge Distribution.}  
Because Tor's list of relays is publicly accessible, Internet blocking authorities can easily discover and block access 
to Tor by IP and port, or through flow-based filtering on easily-fingerprintable traffic characteristics such as fixed
cell size and non-standard TLS configurations.  To counteract these threats, Tor introduced unlisted relays known as {\em bridges}~\cite{dingledine2006design}, and can use a variety of ``pluggable transports'' to evade flow-based filtering~\cite{fifield2015blocking,obfs4,snowflake,webtunnel}.  

However, two remaining problems hinder the ability of Tor to counteract blocking.  First, the signaling problem~\cite{vines2024communication} must be solved, allowing users to communicate with a {\em bridge authority} 
to obtain contact information for these unlisted bridges.  While Tor currently supports a variety of methods, including
an email responder, several CDN web services, manual installation, and Telegram bot, development of other methods remains an active and important area of work. A final problem, as observed by Dingledine and Mathewson~\cite{dingledine2006design,Dingledine11}, is how to balance availability of bridges against the possibility of insider enumeration.  Starting with Proximax~\cite{proximax}, a series of papers~\cite{proximax,wang2013rbridge,douglas2016salmon,nasr2019enemy,tulloch2023lox} have investigated methods of limiting this distribution through trusted ``invitation''-based systems.  Our work is most closely related to three representative systems: rBridge~\cite{wang2013rbridge}, Salmon~\cite{douglas2016salmon}, and Lox~\cite{tulloch2023lox}. We summarize their core mechanisms, clarify the adversarial models they target, and identify the remaining gaps that motivate our design. We also discuss related work on private information retrieval (PIR), which provides one of the central cryptographic building blocks in our construction.

\paragraph{rBridge.}
rBridge proposes a reputation-based approach to bridge distribution: instead of revealing many bridges to every new account, it grants access progressively as a user demonstrates benign behavior over time. The design is motivated by the insider-enumeration problem, where a blocking authority can create many accounts, request bridges, and block them at scale. To mitigate this, rBridge ties bridge allocation to a notion of accumulated \emph{reputation} through repeated interactions, paired with referral-style growth (in which high-reputation users can issue invitations to new users) to enable onboarding without globally exposing bridges~\cite{wang2013rbridge}. rBridge also crystallizes the core availability–security tension: the distributor must keep bridges usable for honest clients while limiting the adversary’s ability to discover and burn the bridge pool. At the same time, it highlights a problem with data-minimized systems: because invitations and bridge assignments are unlinkable across interactions, blocking authorities can use "sock-puppet" invitations to learn and burn bridges while accumulating reputation with unblocked-but-overloaded bridges.

\paragraph{Salmon.}
Salmon studies robust proxy distribution against a network information controller that can create many identities, repeatedly request proxies, and block any proxy it learns~\cite{douglas2016salmon}. Its core design maintains a per-user suspicion score and bans users whose suspicion exceeds a threshold. In parallel, it uses discrete trust levels to control which proxies a user is eligible to receive and how quickly access expands over time. To support onboarding without immediately exposing high-quality proxies, Salmon introduces recommendation-based admission. It records a social graph of peer-to-peer recommendations, and, when possible, assigns users within the same recommendation component to the same proxy servers. This perspective is useful for our setting because it makes explicit that enumeration resistance requires adaptive, stateful policies rather than one-shot disclosure. At the same time, these policies require the distributor to maintain detailed user histories, assignments, and recommendation structures, which creates a metadata-visibility risk that later systems aim to mitigate.

\paragraph{Lox.}
Lox is a modern bridge distribution system that targets both enumeration resistance and user privacy by enabling policy enforcement without requiring persistent user identifiers at the distributor~\cite{tulloch2023lox}. Lox uses unlinkable multi-show anonymous credentials that store an identifier $\Phi$, {\em trust level} $L \in \{0,\dots,4\}$ {\em issuance time} $t$, {\em bridge bucket}, {\em invitation count} $a$, and {\em migration count} $d$.  Users can repeatedly prove authorization to access bridges in their assigned ``bucket'', and depending on their trust level, can request to ``Level Up'' (after a required waiting period), issue a new Invitation to a friend, or ``migrate'' to a new bridge bucket if all currrent bridges are blocked;  these latter requests result in updated credentials.
In this way,  a user can repeatedly prove authorization and satisfy policy checks, e.g., rate limits, eligibility, and trust-level progression, while keeping successive requests cryptographically unlinkable using an algebraic MAC~\cite{chase2014algebraic}. This lets the distributor enforce long-term policies—such as gradually granting access to more reliable users—without learning a stable account identity or reconstructing a user's history.

Lox thus adopts a trust-level perspective similar in spirit to Salmon and reputation-driven systems like rBridge, but instantiates it with cryptographic credentials that reduce metadata leakage about users and their relationships. Lox also highlights privacy goals specific to bridge distribution, including limiting what the distributor can infer about a user's long-term behavior and preventing trivial correlations between bridge assignments and particular users.  However, Lox does not have the ability to adaptively balance bridge load, detect changes in transport filtering, or respond to regional blocking of proxies.

\subsection{Background in Function Secret Sharing.}
\label{subsec:related-fss}

A central primitive in our design is \emph{function secret sharing} (FSS), introduced by Boyle, Gilboa, and Ishai as a way to split a function into short keys such that each server, given only its own key, learns nothing about the function, while the parties' evaluations combine to recover the function value~\cite{boyle2015function}. An important special case is the \emph{distributed point function} (DPF), formalized by Gilboa and Ishai, where the shared function is zero everywhere except at one target point~\cite{gilboa2014distributed}. This point-function view is natural for private lookups and sparse updates, and it is the main FSS flavor used in G-Lox.

Boyle \emph{et al.}\ later improved and extended FSS, giving more efficient constructions and broadening the supported function families, with explicit motivation from applications such as private reads and writes to distributed databases~\cite{boyle2016function}. For our purposes, these works provide the right abstraction for hiding which user, group, or directory entry is being touched, while keeping the online work low and naturally compatible with a two-server architecture.

\paragraph{FSS-backed secret-shared state.}
Our first use of FSS is in maintaining user and group state across two non-colluding servers. Conceptually, the system state is stored in secret-shared form, and updates are expressed through compact FSS keys that describe sparse changes to the logical database. This viewpoint matches the original motivation of FSS for securely searching and updating distributed data~\cite{boyle2015function}. In G-Lox, this allows the servers to apply state transitions without learning which logical record was updated, thereby helping realize the privacy wall between user activity and bridge-assignment logic. More broadly, the line of work on secure computation with preprocessing via FSS shows that FSS is not only a tool for private retrieval, but also a useful building block for efficient two-party computation with acceptable online cost~\cite{boyle2019secure}. 

\paragraph{FSS-based two-server PIR.}
Our second use of FSS is for efficient private directory lookup. Here, the relevant primitive is essentially DPF-based two-server PIR: the client secret-shares a point query into two compact keys, each server evaluates its key over the replicated database, and the client combines the answers to recover exactly one record while each server individually learns nothing about the queried index~\cite{gilboa2014distributed,boyle2016function}. The appeal of this approach in G-Lox is that it fits the system architecture directly: we already assume two non-colluding servers, and the reconstruction rule is simple, while the privacy guarantee is precisely index hiding against either server alone. One server PIR is not suitable for our setting, and it is generally expensive~\cite{colombo2023authenticated, LinMW23_DEPIR, OkadaPlayerPohmannWeinert2024, OkadaPlayerPohmannWeinert2025}.

\section{Cryptographic Preliminaries}
\label{sec:crypto-primitives}
This section fixes the primitives and conventions used throughout G--Lox.
Our design follows the algebraic MAC--based keyed-verification anonymous credential (KVAC)
construction of~\cite{chase2014algebraic}, and augments it with:
(i) pseudorandom functions (PRFs) to derive hidden group tags, deterministic assignment indices,
and deduplication nullifiers;
(ii) authenticated encryption (AEAD-style) for capability objects returned by the back end;
(iii) standard zero-knowledge proofs of knowledge to show well-formed KVACs and required statements; and
(iv) a two-server back end that stores all group-keyed state and executes stateful adaptation behind a
privacy wall.
Concretely, the privacy wall combines a two-server Distributed Point Function and Function Secret Sharing point access for the group-state map with a token-gated two-server DPF-PIR protocol for directory redemption, while state updates are executed via semi-honest Yao two-party computation between the servers. This design keeps the public-facing distributor oblivious to group linkability and bridge assignments.

\subsection{Notation and Hardness Assumptions}
\label{subsec:crypto-notation}

Let $\lambda$ be the security parameter.
We write $x \sample \mathcal{X}$ for sampling $x$ uniformly at random from a finite set $\mathcal{X}$.
For an integer $n$, $[n]:=\{1,\dots,n\}$.
We fix a cyclic group $\mathbb{G}$ of prime order $q$ with generator $g_{gen}$, where discrete logarithms are hard. Concretely, $\mathbb{G}$ can be instantiated as a prime-order elliptic-curve group with $q\approx 2^\lambda$. We use hash functions modeled as random oracles:
\begin{itemize}
  \item $H_{\mathsf{attr}}:\{0,1\}^\ast\to \mathbb{Z}_q$ maps attributes into exponents;
  \item $H_{\mathsf{split}}:\{0,1\}^\ast\to \{0,1\}^{\ell_{\mathsf{split}}}$ for split derivations.
\end{itemize}
We use PRFs $F_{\mathsf{grp}},F_{\mathsf{idx}},F_{\mathsf{init}},F_{\mathsf{dedup}}$ with independent keys.
\begin{itemize}
  \item $F_{\mathsf{grp}}:\{0,1\}^\lambda\times \{0,1\}^\ast\to \{0,1\}^{\ell_{\mathsf{tag}}}$
  derives hidden group tags.
  \item $F_{\mathsf{idx}}:\{0,1\}^\lambda\times \{0,1\}^\ast\to \{0,1\}^{\ell_{\mathsf{idx}}}$
  derives deterministic assignment material.
  \item $F_{\mathsf{init}}:\{0,1\}^\lambda\times \{0,1\}^\ast\to \{0,1\}^{\lambda}$
  derives an initial group label from the invitation material.
  \item $F_{\mathsf{dedup}}:\{0,1\}^\lambda\times \{0,1\}^\ast\to \{0,1\}^{\ell_{\mathsf{nf}}}$
  derives deduplication nullifiers to prevent double-counting.
\end{itemize}
All PRF keys used for group state ($k_{\mathsf{grp}},k_{\mathsf{idx}},k_{\mathsf{init}},k_{\mathsf{dedup}}$) are
secret-shared across the two state servers and evaluated only inside 2PC. 

We use an AEAD scheme $\mathsf{AEAD}=(\mathsf{Enc},\mathsf{Dec})$ with explicit nonces, standard correctness, and IND-CCA-style confidentiality and integrity.
We also use end-to-end public-key encryption to return the user's private outputs.
Concretely, we treat $\mathsf{PKE}=(\mathsf{KeyGen},\mathsf{Enc},\mathsf{Dec})$ as a Key Encapsulation Mechanism-- Data Encapsulation Mechanism (KEM--DEM) instantiation: an IND-CCA secure KEM encapsulates a fresh session key $K$ to the user’s ephemeral $pk_U$,
and an AEAD DEM encrypts the response payload under $K$ similar to techniques in~\cite{cremers2024keeping,len2021partitioning}.

We assume standard hardness of DL/CDH in $\mathbb{G}$, PRF security for the PRFs above, and standard security for AEAD and $\mathsf{PKE}$.

\subsection{Algebraic MAC--based KVAC}
\label{subsec:kvac}

We use the CMZ14 algebraic-MAC KVAC construction~\cite{chase2014algebraic} to encode Lox and G--Lox attributes.
A credential contains an attribute vector
\[
  \mathbf{x}=(x_1,\dots,x_\ell)\in \mathcal{X}_1\times\cdots\times \mathcal{X}_\ell.
\]
Each attribute is mapped injectively into $\mathbb{Z}_q$ via a domain-separated attribute hash
\[
  \widehat{x}_j := H_{\mathsf{attr}}(j \,\|\, \mathrm{encode}(x_j)) \in \mathbb{Z}_q,
\]
which fixes a canonical field representation and prevents cross-attribute collisions.
At a high level, the issuer computes an algebraic MAC on the vector $(\widehat{x}_1,\dots,\widehat{x}_\ell)$ under a secret key,
and the user stores the resulting credential in a form that supports selective disclosure and unlinkable showing: during a show, the user produces a proof that it holds a valid MAC on some attribute vector, optionally revealing a subset of attributes and proving predicates over the rest in zero knowledge, without revealing the hidden attributes.

In G--Lox, we use only \emph{one-show} presentation: each show reveals only the ephemeral identifier $\Phi$, which is used for one-time deduplication at $D$, while sensitive attributes, in particular the hidden group label $\gamma$ and stable hidden identifier $u$, remain hidden and appear only as part of ZK witnesses.
This enables $D$ to enforce policy while remaining blind to group membership and inter-show linkage beyond the
single revealed $\Phi$. The distributor enforces one-show by rejecting any reuse of a previously revealed $\Phi$.

\subsection{Two-Server MPC and Private State Access}
\label{sec:mpc-fss}

To prevent the public distributor $D$ from learning group linkability, all group-keyed state
(epoch, bad-type flags, blockage counters, dedup filters, split seed, size estimate, etc.)
is maintained behind a two-server privacy wall.

\paragraph{Two-party MPC model (2PC).}
Two non-colluding state servers $S_0,S_1$ jointly execute a standard secure two-party computation (2PC)
protocol on secret-shared values.
For bitstrings we use XOR-sharing $x=\langle x\rangle_0\oplus \langle x\rangle_1$, and for arithmetic values we use
additive sharing modulo an appropriate modulus.
Inside 2PC, $S_0,S_1$ evaluate a fixed program $\Pi_{\mathsf{state}}$ implementing bridge assignment,
blockage handling, deduplication, and splitting logic on shares.
We assume the standard simulation-based privacy and correctness notion for 2PC in the semi-honest model. Our circuits can be instantiated using either Yao garbled circuits or the Goldreich--Micali--Wigderson (GMW) protocol~\cite{yao1986generate,goldreich2019play} and We primarily choose Yao’s protocol because it evaluates a Boolean circuit in a constant number of online rounds, with inputs supplied via standard oblivious transfer implemented using OT extension. Online communication is then dominated by the non-linear gates, making Yao attractive when latency is non-negligible or when circuits have large Boolean depth. By contrast, GMW requires online interaction proportional to circuit depth, but can be competitive in very low-latency settings and under heavy batching of many independent instances.

\paragraph{FSS and DPF.}
A two-server distributed point function is a function secret sharing scheme for point functions.
For a domain $[M]=\{0,\dots,M-1\}$, $\mathsf{DPF.Gen}(j)$ outputs two short keys $(k_0,k_1)$ such that, for every $\ell\in[M]$,
\[
\mathsf{DPF.Eval}(k_0,\ell)\ \oplus\ \mathsf{DPF.Eval}(k_1,\ell)\;=\;\mathbf{1}\{\ell=j\},
\]
and any single key $k_b$ computationally hides $j$. Intuitively, each server holds an additive share of the length-$M$ indicator vector that selects exactly one position.

For the group-state map keyed by $\mathsf{tag}$, we implement private point access using two-server FSS for point functions,
instantiated by DPF. Concretely, view the map as an array of $M$ fixed-size records $\mathsf{DB}[0..M{-}1]$, where each record
is $B_{\mathsf{state}}$ bits, packed into $B=B_{\mathsf{state}}/8$ bytes, and the index $j$ is derived from $\mathsf{tag}$.
To read address $j$, the client generates two DPF keys $(k_0,k_1)\gets \mathsf{DPF.Gen}(j)$ and sends $k_b$ to server $S_b$.
Each server locally expands its key and returns an XOR-share of the record:
\[
y_b \;:=\; \bigoplus_{\ell=0}^{M-1} \mathsf{DB}[\ell]\cdot \mathsf{DPF.Eval}(k_b,\ell)\ \in\ \{0,1\}^{B_{\mathsf{state}}},
\]
so that $y_0\oplus y_1=\mathsf{DB}[j]$, while neither server learns $j$ under non-collusion. We denote this interface by
$\mathsf{FSS.Read}(\mathsf{tag})$, which returns XOR-shares of the $B_{\mathsf{state}}$-bit record.

For updates, we use the standard point-update variant. To apply a masked delta $\Delta\in\{0,1\}^{B_{\mathsf{state}}}$ to address $j$,
the privacy wall generates keys for the point function $\ell\mapsto \Delta\cdot \mathbf{1}\{\ell=j\}$ and sends them to $S_0,S_1$.
Each server XORs its expanded contribution into its local database, yielding a logical operation
$\mathsf{FSS.Write}(\mathsf{tag},\Delta)$ that updates $\mathsf{DB}[j]\leftarrow \mathsf{DB}[j]\oplus \Delta$ without revealing $j$.
This provides private key--value access under the non-collusion assumption and avoids scan-and-reshuffle for the group map.
When the state grows beyond fixed-size records, or when we must hide richer access patterns, we can instead place the group map behind
an ORAM-backed array to obtain $O(\log M)$ bandwidth per logical access at the cost of a larger 2PC circuit~\cite{doerner2017scaling, chen2019onion,stefanov2018path, doerner2017scaling}.

\section{G-Lox: Group-Adaptive Lox}
\label{sec:glox}

G--Lox builds on the Lox interface and preserves its front-end credential workflow:
users obtain CMZ14-style credentials from the distributor, present one-show proofs,
and evolve through Lox-style trust levels.  However, differently to Lox, user credentials
in G--Lox retain a hidden group identifier that groups together users that have a common
ancestor in the invitation tree.  Because we expect users to primarily issue invitations 
to other users in the same region, this allows G--Lox to respond adaptively to region-specific
changes in network information controls, privately accumulating evidence of new transport filtering
or blockage, while also further limiting the effect of Sybil and network-level surveillance attacks.

Relative to Lox, the main algorithmic change is that bridge distribution becomes \emph{stateful} and \emph{group-adaptive}: group-keyed
state and bridge-selection logic are moved behind a two-server ``privacy wall'', so the public-facing distributor remains oblivious to group membership and bridge
assignments. We therefore focus in this section on the new stateful and
privacy-preserving components, and treat the inherited Lox issuance and showing machinery as unchanged unless explicitly noted; see Appendix~\ref{app: low_review}
or~\cite{tulloch2023lox} for further background on Lox.

Concretely, two non-colluding state servers $S_0, S_1$ maintain the group-state map as a two-server DPF/FSS-backed key--value store and secret-share all long-term symmetric keys.
On each \textsf{GetBridge} request, they run a small semi-honest Yao 2PC to derive the hidden group tag
$\mathsf{tag}:=F_{\mathsf{grp}}(k_{\mathsf{grp}},\gamma)$, privately read/update the $\mathsf{tag}$-indexed state record,
and compute the bridge assignment from PRFs on $(\mathsf{tag},e,\tau)$ for stable rotation and consistent adaptation. The public distributor $D$ is purely front-end: it verifies one-show proofs, enforces non-reuse of the revealed
identifier $\Phi$, and relays messages, while all linkable outputs are returned only as end-to-end ciphertexts to the client. Bridge redemption is handled by a directory that partitions $\mathsf{Dir}$ by transport type $\tau$ and supports token-gated lookup via two-server DPF-PIR, matching the same two-server trust split.

Before presenting the construction, we recall our assumptions and threat model.

\paragraph{Threat model.}
We consider an active blocking authority that can observe and block public entry points, mount Sybil-style enumeration to learn and burn bridges, and selectively interfere with connections in ways that may be vantage-dependent, partial, and transient. We further assume the action of information control is often transport-specific and region/group-specific, so blocking pressure may vary across transports and user groups~\cite{bhaskar2024understanding}.
On the infrastructure side, the public-facing distributor $D$ may be honest-but-curious and must learn neither users' bridge assignments nor their group membership/linkability across requests. Our privacy guarantees rely on two non-colluding state servers $S_0, S_1$: as long as they do not collude, MPC+FSS hides group-keyed state and access patterns; if they collude, these privacy protections fail.

\subsection{Settings}
\label{subsec:glox-settings}

\paragraph{Participants.}
The system comprises a public-facing distributor $D$ that verifies and issues G--Lox credentials via CMZ14-style
KVAC showing/issuance and relays protocol messages; two non-colluding state servers $S_0, S_1$ that hold secret shares of all long-term symmetric keys, maintain group-state databases jointly via 2PC, and hold a bridge dictionary~\cite{van2010bridgedb} where the users can redeem their bridge token privately; and a set of users who interact with $D$. An anonymized connection to $D$ is not required for protocol correctness, however without it use-frequency metadata can leak.  Our privacy guarantees are stated at the application layer: even without network-layer anonymity, $D$ should not learn hidden group membership or bridge assignments from the protocol transcript.

\paragraph{System parameters.}
We fix security parameter $\lambda$ and transport-type set $\mathcal{T}$; a per-group report threshold $\theta$ that triggers migration; a group-size bound $G_{\max}$ that triggers splitting; a split-transition window $W\in\mathbb{N}$ during which the parent group may remain active for stragglers; Lox-style policy arrays
$\textsf{DAYS}[1..4]$, $\textsf{INVITATIONS}[1..4]$, $\textsf{MAX\_BLOCK}[1..4]$, and $\textsf{MIN\_REP}[1..4]$; and secret-shared keys held by $S_0,S_1$ and used only inside 2PC:
$\langle k_{\mathsf{grp}}\rangle$, $\langle k_{\mathsf{idx}}\rangle$, $\langle k_{\mathsf{tok}}\rangle$,
$\langle k_{\mathsf{init}}\rangle$, $\langle k_{\mathsf{pir}}\rangle$, and $\langle k_{\mathsf{dedup}}\rangle$. Finally, each request carries a fresh user public key $pk_U$ for end-to-end encryption under $\mathsf{PKE}$.

\paragraph{Per-group state components.}
Each group-state record contains several compact fields used by the back-end policy logic. First, \emph{bad-type bits} are a bit-vector over transport types in $\mathcal T$; the bit for $\tau\in\mathcal T$ is set when the system has accumulated sufficient evidence that this
transport is currently unsuitable for the group, so future assignments for that group/epoch should avoid $\tau$. Second, a \emph{dedup filter} stores short fingerprints of recently processed reports so that repeated submissions of the same effective event do not inflate
the corresponding per-group counters. Concretely, in \textsf{ReportBlocked} we derive a fingerprint such as $F_{\mathsf{dedup}}(k_{\mathsf{dedup}},u\|\tau\|e\|i)$ and count the report only if this fingerprint is fresh. These structures are maintained inside the FSS-backed group-state map and are never revealed to the public distributor.

\paragraph{Databases.}
(1) A group-state map indexed by $\mathsf{tag}:=F_{\mathsf{grp}}(k_{\mathsf{grp}},\gamma)$, implemented using
two-server FSS/DPF (\textsf{FSS.Read}/\textsf{FSS.Write}) over fixed-size records that store epoch, bad-type bits,
small counters, dedup filters, and optional split seed/state.
(2) A bridge directory that two state servers both hold, and the user retrieves the bridge via two-party FSS-based efficient PIR.  

\paragraph{G-Lox credential.}
We extend the Lox credential with group and behavior attributes:
\[
  \Psi^{\mathsf{G}}=(\Phi,t,L,a,d,\gamma,r,u),
\]
where \(\Phi\) is the token identifier used to prevent reuse; \(t\) is the join-time attribute used for time-based trust evolution; \(L\) is the user's trust level; \(a\) is the available-invitations counter; and \(d\) is a blockage/migration counter used to bound eligibility under repeated blocking. The group label $\gamma\in\{0,1\}^\lambda$ is a uniformly random value shared by users in the same hidden group and is
always hidden from $D$ such that it appears only as a witness in ZK proofs. The value $r$ is a reputation counter checked in the level-up proof. In our design, $r$ is
revealed only during \textsf{ReportBlocked} so that $D$ can update it, while $\gamma$ and $u$ remain hidden.
Specifically, $r$ is updated only during credential re-issuance based on a one-bit outcome
$\mathsf{contrib}\in\{0,1\}$ produced by 2PC (Algorithm~\ref{alg:glox-reportblocked}).
Finally, $u$ is a stable, high-entropy hidden identifier used only inside 2PC logic, never revealed to $D$.

\paragraph{Group label forming and Type Selection.}
To instantiate hidden groups, users receive $\gamma$ at issuance derived from invitation material.
During issuance/update, the user supplies an invitation token $\mathsf{invTok}$ to the state layer, and the servers compute
\[
  \gamma := F_{\mathsf{init}}(k_{\mathsf{init}},\mathsf{invTok})
\]
under secret-shared $\langle k_{\mathsf{init}}\rangle$; the resulting $\gamma$ is issued as a hidden attribute.
Users presenting the same $\mathsf{invTok}$ receive the same hidden $\gamma$ and thus form a hidden group. Per~\cite {tulloch2023lox}, such a group can be used to construct the users' social graph, so we shall not reveal it.

To choose the transport type $\tau$ deterministically, two servers derive a pseudorandom selector
$u \gets F_{\mathsf{idx}}(k_{\mathsf{idx}},\,\mathsf{tag}\,\|\,\texttt{"TYPE"})$
and run $\tau \leftarrow \textsf{SelectType}(u,\mathsf{st.badTypes})$, which maps $u$ to a type in the allowed set. Thus $\tau$ is consistent under replays and automatically avoids types that the group has marked as blocked.

\paragraph{Bridge token.}
In \textsf{GetBridge}, the MPC state layer outputs an \emph{opaque bridge token} $\Omega^{\mathsf B}$ that binds the issued
assignment to a later \textsf{ReportBlocked} update, while keeping the directory logic from having to interpret any
group-derived identifiers. Concretely, inside 2PC, the servers derive
$\mathsf{tag}:=F_{\mathsf{grp}}(k_{\mathsf{grp}},\gamma)$, read group state to obtain the current epoch $e$,
choose a transport type $\tau$, compute the directory index $i$,
sample a fresh nonce $\nu\sample\{0,1\}^\lambda$, and form
\[
\Omega^{\mathsf B}
:=\Bigl(\nu,\ \mathsf{AEAD.Enc}_{k_{\mathsf{tok}}}\bigl(\nu;\ \mathsf{AD}=\texttt{"BRIDGE"},\ \mathsf{msg}=(\tau,e,i)\bigr)\Bigr).
\]
The associated-data label \texttt{"BRIDGE"} is a fixed, system-wide constant (independent of $\mathsf{tag}$), so the token format does not encode any hidden group identifier. The client receives $\Omega^{\mathsf B}$ but treats it as an opaque
string: only the MPC backend, holding $k_{\mathsf{tok}}$ in secret-shared form, can later open it inside 2PC.

We also include a one-time PIR authorization token. Inside 2PC, the servers sample a
fresh $\eta\sample\{0,1\}^\lambda$ and set $\mathsf{exp}\leftarrow \mathsf{now}+\Delta_{\mathsf{pir}}$,
then compute
\[
\sigma := \mathsf{MAC}_{k_{\mathsf{pir}}}\bigl(\texttt{"PIR"}\|\eta\|\tau\|\mathsf{exp}\bigr),
\qquad
\Omega^{\mathsf{pir}} := (\eta,\tau,\mathsf{exp},\sigma).
\]
The client presents $\Omega^{\mathsf{pir}}$ to the directory to authorize exactly one subsequent \textsf{DirPIR} query.

\subsection{G-Lox Protocols}
In this subsection, we concretely present our core construction of the G-Lox protocol. We reuse Lox Constructions~\ref{alg:lox-open-entry}--\ref{alg:lox-redeem} with the following modifications:
every issuance or update extends the attribute vector with hidden $\gamma$, $r$, and $u$.
We remove the bucket-reachability credential $\Psi^{\mathsf R}$, and the level-up protocol checks behavior directly
via $d$ and $r$. We present the modified level-up protocol, then the bridge-related protocols.

\paragraph{Level up.}
The G--Lox level-up protocol upgrades a user’s trust from $L$ to $L{+}1$ (for $1\le L<4$) without revealing sensitive
attributes or enabling linkage across shows as shown in Algorithm~\ref{alg:glox-levelup}. The user reveals the current one-show identifier $\Phi$ and $L$, and proves
in zero knowledge that the hidden attributes satisfy the policy predicates (time, $d$, and $r$).
To ensure unlinkability, the user and $D$ jointly derive a fresh identifier $\Phi'$ and $D$ issues the new credential under $\Phi'$, while enforcing non-reuse of $\Phi$.

\begin{algorithm}[t]
\caption{\textsf{G-Lox LevelUp} (trust $L \to L{+}1$; rotate one-show identifier)}
\label{alg:glox-levelup}
\begin{algorithmic}[1]
\Require User holds $\Psi^{\mathsf G}=(\Phi,t,L,a,d,\gamma,r,u)$ with $1\le L<4$.
\Require Distributor $D$ holds policy arrays \textsf{DAYS}, \textsf{MAX\_BLOCK}, \textsf{MIN\_REP}.
\Ensure User obtains $\Psi^{\mathsf G}_{\mathsf{new}}=(\Phi',t',L{+}1,a',d,\gamma,r,u)$.

\Statex \textbf{User:}
\State Sample fresh $\Phi' \sample \{0,1\}^{\lambda}$
\State $\pi_{\mathsf{lvl}}\gets \mathsf{ZKProve}[\ \text{reveal }\Phi,L;\ \text{hide }(t,a,d,\gamma,r,u);\ \text{valid one-show};$
\Statex \hspace{2.2em} $t+\textsf{DAYS}[L{+}1]\le \textsf{now}\ \wedge\ d\le \textsf{MAX\_BLOCK}[L{+}1]\ \wedge\ r\ge \textsf{MIN\_REP}[L{+}1]\ ]$
\State Send \textsf{LevelUpReq}$\langle \Phi,L,\Phi',\pi_{\mathsf{lvl}}\rangle$ to $D$

\Statex \textbf{Distributor $D$:}
\State Verify $\pi_{\mathsf{lvl}}$; reject if $\Phi$ was previously used
\State Set $t'\gets \textsf{now}$; set $a'\gets \textsf{INVITATIONS}[L{+}1]$
\State Issue new credential $\Psi^{\mathsf G}_{\mathsf{new}}=(\Phi',t',L{+}1,a',d,\gamma,r,u)$ via CMZ14 issuance
\State Return \textsf{LevelUpResp}$\langle \Psi^{\mathsf G}_{\mathsf{new}}\rangle$ to user
\end{algorithmic}
\end{algorithm}

\paragraph{Get Bridge and Redeem Bridge}
Algorithm~\ref{alg:glox-getbridge-fss} describes the end-to-end \textsf{GetBridge} flow with (i) an FSS/DPF-backed \emph{group-state} map and (ii) a \emph{PIR-gated} directory redemption. The user first proves, in zero knowledge, possession of a valid one-show credential while revealing only the show identifier $\Phi$, which $D$ uses to enforce one-show and rate limits. The distributor $D$ then acts only as a relay into a 2PC between $(S_0, S_1)$: the servers derive the hidden group tag $\mathsf{tag}=F_{\mathsf{grp}}(k_{\mathsf{grp}},\gamma)$, privately read/update the corresponding group state via \textsf{FSS.Read}/\textsf{FSS.Write}, and deterministically sample an assignment $(\tau, i)$ for the current epoch $e$. They return the following to the user via an encrypted channel under $pk_U$: (a) an opaque bridge token $\Omega^{\mathsf B}$ that later authenticates \textsf{ReportBlocked} inside 2PC, and (b) a short-lived, one-time authorization token $\Omega^{\mathsf{pir}}$ that permits exactly one directory query for transport type $\tau$.

Algorithm~\ref{alg:glox-redeemdirpir} shows how the user redeems $\Omega^{\mathsf{pir}}$ to learn the actual bridge descriptor $\mathsf{Dir}_\tau[i]$ using two-server DPF-PIR. The user generates DPF keys $(k_0,k_1)$ for index $i$ and sends one key to each non-colluding directory server along with $\Omega^{\mathsf{pir}}$. Each server locally checks freshness and MAC validity, enforces one-time use via a spent-set on $\eta$, and answers with its PIR share $y_b$ computed by evaluating the DPF over all indices and XOR-aggregating the selected record. The user reconstructs $\mathsf{bridgeDesc}$ as $y_0\oplus y_1$. Under non-collusion, neither server learns $i$, while the token gate prevents unauthorized or replayed directory queries.

\begin{algorithm}[htbp!]
\caption{\textsf{G-Lox GetBridge} (FSS-backed group state + PIR authorization)}
\label{alg:glox-getbridge-fss}
\begin{algorithmic}[1]
\Require User holds $\Psi^{\mathsf G}=(\Phi,t,L,a,d,\gamma,r,u)$ with $L\ge 1$.
\Require Distributor $D$ verifies KVAC shows and enforces one-show by rejecting reuse of $\Phi$.
\Require $(S_0,S_1)$ holds an FSS/DPF-backed group-state map and secret-shared keys
$\langle k_{\mathsf{grp}}\rangle,\langle k_{\mathsf{idx}}\rangle,\langle k_{\mathsf{tok}}\rangle,\langle k_{\mathsf{pir}}\rangle$.
\Ensure User obtains $(\tau,i,e,\Omega^{\mathsf B},\Omega^{\mathsf{pir}},\mathsf{SplitInfo})$ without revealing $\mathsf{tag}$ or $(\tau,i)$ to $D$.

\Statex \textbf{User $\to D$: show + relay}
\State $(pk_U,sk_U)\gets \mathsf{KeyGen}(1^\lambda)$
\State $\pi_{\mathsf{cred}}\gets \mathsf{ZKProve}[\ \text{reveal }\Phi;\ \text{hide }(t,\cdots,u);\ \text{valid one-show};\ L\ge 1;\ \text{binds }(\gamma,u)\ ]$
\State Send \textsf{GetBridgeReq}$\langle \Phi,pk_U,\pi_{\mathsf{cred}}\rangle$ to $D$; provide MPC shares of $\gamma$

\Statex \textbf{$D$: verify and forward}
\State Verify $\pi_{\mathsf{cred}}$; reject if $\Phi$ was previously used
\State Forward request + $\gamma$ shares to $(S_0,S_1)$

\Statex \textbf{2PC on $(S_0,S_1)$: private state access + token generation}
\State $\mathsf{tag}\gets F_{\mathsf{grp}}(k_{\mathsf{grp}},\gamma)$
\State $\mathsf{st}\gets \textsf{FSS.Read}(\mathsf{tag})$ \Comment{$\mathsf{st}$ holds epoch $e$, $\mathsf{badTypes}$, etc.}
\State $e\gets \mathsf{st}.\mathrm{epoch}$
\State $\tau \gets \mathsf{SelectType}(F_{\mathsf{idx}}(k_{\mathsf{idx}},\mathsf{tag}\|e\|\texttt{"TYPE"}),\,\mathsf{st}.\mathrm{badTypes})$
\State $i \gets F_{\mathsf{idx}}(k_{\mathsf{idx}},\mathsf{tag}\|e\|\tau)\bmod N_\tau$

\State Sample $\nu \sample \{0,1\}^{\lambda}$
\State $\Omega^{\mathsf B}\gets \bigl(\nu,\ \mathsf{AEAD.Enc}_{k_{\mathsf{tok}}}(\nu;\ \textsf{AD}=\texttt{"BRIDGE"},\ (\tau,e,i))\bigr)$

\State Sample $\eta \sample \{0,1\}^{\lambda}$; set $\textsf{exp}\gets \textsf{now}+\Delta_{\mathsf{pir}}$
\State $\sigma \gets \mathsf{MAC}_{k_{\mathsf{pir}}}\bigl(\texttt{"PIR"}\|\eta\|\tau\|\textsf{exp}\bigr)$
\State $\Omega^{\mathsf{pir}} \gets (\eta,\tau,\textsf{exp},\sigma)$ \Comment{one-time authorization for one DirPIR query}

\State Update $\mathsf{st}$ as needed; $\textsf{FSS.Write}(\mathsf{tag},\Delta_{\mathsf{st}})$
\State $C \gets \mathsf{Enc}_{pk_U}\big(\tau,i,e,\Omega^{\mathsf B},\Omega^{\mathsf{pir}},\mathsf{SplitInfo}\big)$
\State Output $C$ to $D$

\Statex \textbf{$D$: respond}
\State Re-issue fresh one-show identifier $\Phi'$ (standard Lox show semantics) and return $C$ to user
\end{algorithmic}
\end{algorithm}

\begin{algorithm}[htbp!]
\caption{\textsf{G-Lox RedeemDirPIR} (token-gated two-server DPF-PIR for $\mathsf{bridgeDesc}$)}
\label{alg:glox-redeemdirpir}
\begin{algorithmic}[1]
\Require Two non-colluding directory servers $S_0,S_1$ each store the same directory partition
$\mathsf{Dir}_\tau[0..N_\tau{-}1]$ of fixed-length descriptors.
\Require Each server maintains a local spent-set $\mathsf{Spent}_b$ of previously redeemed nonces $\eta$.
\Require User holds $(\tau,i,e,\Omega^{\mathsf{pir}})$ from \textsf{GetBridge}, where
$\Omega^{\mathsf{pir}}=(\eta,\tau,\textsf{exp},\sigma)$ and $\sigma=\mathsf{MAC}_{k_{\mathsf{pir}}}(\texttt{"PIR"}\|\eta\|\tau\|\textsf{exp})$.
\Require A DPF/FSS scheme for point functions with algorithms $\mathsf{DPF.Gen},\mathsf{DPF.Eval}$ as in~\cite{boyle2015function}.
\Ensure User learns $\mathsf{bridgeDesc}=\mathsf{Dir}_\tau[i]$; neither $S_0$ nor $S_1$ learns $i$ (under non-collusion);
$\Omega^{\mathsf{pir}}$ is usable at most once.

\Statex \textbf{User: key generation}
\State $(k_0,k_1)\gets \mathsf{DPF.Gen}(i)$

\Statex \textbf{User $\to S_0,S_1$: send query}
\State Send $(\tau,k_0,\Omega^{\mathsf{pir}})$ to $S_0$ and $(\tau,k_1,\Omega^{\mathsf{pir}})$ to $S_1$

\Statex \textbf{Server $S_b$: authorize, answer, and spend token}
\For{$b\in\{0,1\}$ \textbf{in parallel}}
  \State Parse $\Omega^{\mathsf{pir}}=(\eta,\tau,\textsf{exp},\sigma)$
  \State Reject if $\tau$ mismatches, or $\textsf{exp}<\textsf{now}$
  \State Reject unless $\mathsf{VerifyMAC}_{k_{\mathsf{pir}}}\bigl(\sigma;\ \texttt{"PIR"}\|\eta\|\tau\|\textsf{exp}\bigr)=1$
  \State Reject if $\eta\in \mathsf{Spent}_b$; else insert $\eta$ into $\mathsf{Spent}_b$

  \State $y_b \gets 0^{8D'}$\Comment{each element in $\mathsf{Dir}$ is embedded as a $D'$ bytes string}
  \For{$\ell=0$ \textbf{to} $N_\tau-1$}
    \State $s \gets \mathsf{DPF.Eval}(k_b,\ell)$ \Comment{$s\in\{0,1\}$}
    \If{$s=1$}
      \State $y_b \gets y_b \oplus \mathsf{Dir}_\tau[\ell]$
    \EndIf
  \EndFor
  \State Return $y_b$ to the user
\EndFor

\Statex \textbf{User: reconstruction}
\State $\mathsf{bridgeDesc}\gets y_0\oplus y_1$
\State \Return $\mathsf{bridgeDesc}$
\end{algorithmic}
\end{algorithm}

\paragraph{Report Blocked}
Algorithm~\ref{alg:glox-reportblocked} handles a user report while keeping both the hidden group tag and any directory index information private from the distributor. The user first performs a one-show credential presentation that reveals only $(\Phi,r)$, where $\Phi$ enforces one-show and $r$ enables reputation update, and proves in zero knowledge that the request is well-formed and bound to the hidden attributes $(\gamma,u)$. The distributor $D$ verifies the proof, rejects reuse of $\Phi$, and then relays the request to the two state servers, which run a 2PC.

Inside 2PC, $(S_0,S_1)$ derive $\mathsf{tag}=F_{\mathsf{grp}}(k_{\mathsf{grp}},\gamma)$ and privately access the per-group state $\mathsf{st}$ via the FSS-backed map. They then validate the stored bridge token $\Omega^{\mathsf B}$ by AEAD-decrypting it under the secret-shared token key $k_{\mathsf{tok}}$ to recover the assigned tuple $(\tau,e,i)$, and enforce an epoch-consistency rule, i.e., reject unless $e$ matches the current epoch in $\mathsf{st}$. To prevent duplicate counting, the servers compute a deduplication fingerprint $\mathsf{nf}=F_{\mathsf{dedup}}(k_{\mathsf{dedup}},u\|\tau\|e\|i)$ and only increment the counter $\mathsf{st}.\mathrm{ctr}[\tau,i]$ if $\mathsf{nf}$ is fresh. The bit $\mathsf{contrib}$ records whether this report is the threshold-crossing event that raises the counter to $\theta$.

If $\mathsf{contrib}=1$, the state servers perform thresholded migration: they advance the group epoch, sample a fresh assignment $(\tau_{\mathsf{new}}, i_{\mathsf{new}})$ for the new epoch, and generate fresh post-migration material, namely a new opaque bridge token $\Omega^{\mathsf B}_{\mathsf{new}}$ and a one-time, short-lived PIR authorization token $\Omega^{\mathsf{pir}}_{\mathsf{new}}$. Otherwise, these fields are set to $\bot$. The servers commit the updated state back to the FSS store and return an encrypted response $C$ to the user, containing an acknowledgement, the contribution bit that is visible to $D$, and the new assignment material.

Finally, $D$ updates the user’s reputation using $\mathsf{contrib}$, i.e., rewarding threshold-crossing contributions and optionally penalizing non-contributing reports, re-issues the credential with a fresh identifier $\Phi'$ under standard one-show semantics, and forwards the encrypted response. If migration occurred, the user redeems the new assignment by invoking \textsf{RedeemDirPIR} with $\Omega^{\mathsf{pir}}_{\mathsf{new}}$ to fetch the new bridge descriptor and replace the locally stored bridge token.

\begin{algorithm}[t]
\caption{\textsf{G-Lox ReportBlocked}}
\label{alg:glox-reportblocked}
\begin{algorithmic}[1]
\Require $\Psi^{\mathsf G}=(\Phi,t,L,a,d,\gamma,r,u)$ with $L\ge 1$, bridge token $\Omega^{\mathsf B}$.
\Require $D$ enforces one-show use; $(S_0,S_1)$ maintain FSS-backed group state.
\Ensure Updated credential and, if $\mathsf{contrib}=1$, fresh assignment material.

\State $(pk_U,sk_U)\gets \mathsf{KeyGen}(1^\lambda)$
\State $\pi_{\mathsf{cred}}\gets
\mathsf{ZKProve}[
\text{reveal }(\Phi,r);
\text{ hide }(t,\cdots,u);
\text{valid one-show};
L\ge 1;
\text{binds }(\gamma,u)
]$
\State User sends
$\langle \Phi,r,pk_U,\Omega^{\mathsf B},\pi_{\mathsf{cred}}\rangle$
to $D$ and secret-shares $(\gamma,u)$ to $(S_0,S_1)$
\State $D$ rejects unless $\mathsf{ZKVerify}(\pi_{\mathsf{cred}})=1$ and $\Phi$ is fresh

\Statex \textbf{2PC by $(S_0,S_1)$}
\State $\mathsf{tag}\gets F_{\mathsf{grp}}(k_{\mathsf{grp}},\gamma)$
\State $\mathsf{st}\gets \mathsf{FSS.Read}(\mathsf{tag})$
\State $(b,\tau,e,i)\gets \mathsf{OpenBridge}_{k_{\mathsf{tok}}}(\Omega^{\mathsf B})$
\State Reject if $b=0$ or $e\neq \mathsf{st}.\mathrm{epoch}$
\State $\mathsf{nf}\gets F_{\mathsf{dedup}}(k_{\mathsf{dedup}},u\|\tau\|e\|i)$
\State $\mathsf{fresh}\gets \mathsf{Dedup.Insert}(\mathsf{st},\mathsf{nf})$
\State $\mathsf{old}\gets \mathsf{st}.\mathrm{ctr}[\tau,i]$
\State $\mathsf{st}.\mathrm{ctr}[\tau,i]\gets \mathsf{old}+\mathsf{fresh}$
\State $\mathsf{contrib}\gets
\mathbf{1}[\mathsf{fresh}=1\wedge \mathsf{old}+1=\theta]$

\If{$\mathsf{contrib}=1$}
  \State $\mathsf{st}.\mathrm{epoch}\gets \mathsf{st}.\mathrm{epoch}+1$
  \State $(\tau',i')\gets \mathsf{SampleAssign}(k_{\mathsf{idx}},\mathsf{tag},\mathsf{st})$
  \State $\Omega^{\mathsf B}_{\mathsf{new}}\gets
  \mathsf{MakeBridgeTok}_{k_{\mathsf{tok}}}(\tau',\mathsf{st}.\mathrm{epoch},i')$
  \State $\Omega^{\mathsf{pir}}_{\mathsf{new}}\gets
  \mathsf{MakePIRTok}_{k_{\mathsf{pir}}}(\tau')$
\Else
  \State $(\tau',i',\Omega^{\mathsf B}_{\mathsf{new}},\Omega^{\mathsf{pir}}_{\mathsf{new}})
  \gets(\bot,\bot,\bot,\bot)$
\EndIf

\State $\mathsf{FSS.Write}(\mathsf{tag},\mathsf{st})$
\State $C\gets
\mathsf{Enc}_{pk_U}(
1,\mathsf{contrib},\tau',i',
\Omega^{\mathsf B}_{\mathsf{new}},
\Omega^{\mathsf{pir}}_{\mathsf{new}})$
\State Output $C$ to $D$

\Statex \textbf{Return path}
\State $r'\gets [r+\alpha_{\mathsf{rep}}\mathsf{contrib}
-\beta_{\mathsf{rep}}(1-\mathsf{contrib})]_0^{R_{\max}}$
\State $D$ re-issues $\Psi_{\mathsf{new}}^{\mathsf G}$ with fresh $\Phi'$ and reputation $r'$
\State User decrypts $C$; if $\Omega^{\mathsf{pir}}_{\mathsf{new}}\neq\bot$, call
$\textsf{RedeemDirPIR}(\tau',i',\Omega^{\mathsf{pir}}_{\mathsf{new}})$
\end{algorithmic}
\end{algorithm}

\paragraph{Group Split}
Algorithm~\ref{alg:glox-groupsplit} gives a lazy split procedure for oversized hidden groups.
Splitting is triggered inside an ordinary 2PC call: the servers read the hidden group state under
\(\mathsf{tag}=F_{\mathsf{grp}}(k_{\mathsf{grp}},\gamma)\), and if the group exceeds \(G_{\max}\), they sample
a split seed \(s\) and return split metadata only inside the end-to-end encrypted response.  Each user
then migrates locally by computing
\[
b=\mathrm{LSB}(H_{\mathsf{split}}(u\|s)),
\qquad
\gamma'=H_{\mathsf{split}}(\gamma\|s\|b).
\]
Thus branch choice is deterministic for the user but hidden from the distributor and unlinkable across
users.  The routing window \(W\) allows stragglers to keep using the parent tag temporarily; after \(W\)
expires, parent-tag requests return the same split metadata and force migration.

\begin{algorithm}[t]
\caption{\textsf{G-Lox GroupSplit}}
\label{alg:glox-groupsplit}
\begin{algorithmic}[1]
\Require Group state keyed by \(\mathsf{tag}=F_{\mathsf{grp}}(k_{\mathsf{grp}},\gamma)\); threshold \(G_{\max}\); window \(W\).
\Ensure User replaces \(\gamma\) by child secret \(\gamma'\) if split metadata is returned.

\Statex \textbf{2PC by \((S_0,S_1)\)}
\State \(\mathsf{st}\gets \mathsf{FSS.Read}(\mathsf{tag})\)
\If{\(\mathsf{st}.\mathsf{size}>G_{\max}\wedge \mathsf{st}.\mathsf{splitting}=0\)}
  \State \(s\sample\{0,1\}^{\lambda}\)
  \State \(\mathsf{st}.\mathsf{splitting}\gets 1\), \(\mathsf{st}.s\gets s\), \(\mathsf{st}.W\gets W\)
  \State \(\mathsf{FSS.Write}(\mathsf{tag},\mathsf{st})\)
\EndIf
\State \(\mathsf{SplitInfo}\gets(\mathsf{st}.\mathsf{splitting},\mathsf{st}.s,\mathsf{st}.W)\)
\State Include \(\mathsf{SplitInfo}\) in the encrypted response \(C\)

\Statex \textbf{User migration}
\State User decrypts \(C\) and parses \(\mathsf{SplitInfo}\)
\If{\(\mathsf{SplitInfo}.\mathsf{splitting}=1\)}
  \State \(s\gets \mathsf{SplitInfo}.s\)
  \State \(b\gets \mathrm{LSB}(H_{\mathsf{split}}(u\|s))\)
  \State \(\gamma'\gets H_{\mathsf{split}}(\gamma\|s\|b)\)
  \State Store \(\gamma'\) for subsequent credential presentations
\EndIf

\Statex \textbf{Routing}
\State Subsequent calls use secret-shared \(\gamma'\)
\State \(\mathsf{tag}'\gets F_{\mathsf{grp}}(k_{\mathsf{grp}},\gamma')\)
\State Servers route the call to the child state under \(\mathsf{tag}'\)
\State Accept parent \(\mathsf{tag}\) during \(W\); after expiry, return \(\mathsf{SplitInfo}\) and require migration
\end{algorithmic}
\end{algorithm}

\subsection{Complexity}
\label{sec:complexity-getbridge-pir}

We analyze (i) garbled-circuit (GC) cost in AND-gate count, (ii) private group-state access cost for one
read+write of a $B$-byte record under our FSS/DPF backend, (iii) directory retrieval cost via token-gated two-server
DPF-PIR. We fix $\lambda=128$ for GC security and use Yao garbled circuits with
Free-XOR and Half-Gates rules. 

\paragraph{Notation.}
Let $G_{\wedge}(f)$ denote the AND-gate count of the Boolean circuit for function $f$, and the XOR operation is free.
Let $C_{\mathrm{OT}}(n_E)$ denote the amortized online cost for transferring $n_E$ evaluator input bits via OT/OT-extension.
Let $C_{\mathsf{dpfrw}}(M,B),T_{\mathsf{dpfrw}}(M,B)$ be the communication and per-server work for one FSS/DPF
group-state read+write on an array of $M$ logical records of $B$ bytes each.

For the directory, let $\mathsf{Dir}_\tau[0..N_\tau{-}1]$ be the per-transport partition of $N_\tau$ fixed-size
descriptors, each $D$ bytes. Let $|k_{\mathsf{DPF}}(N_\tau)|$ be the size (bytes) of one DPF query key for domain size $N_\tau$. Let $G_{\wedge}^{\mathsf{PRF}}$ be the AND-gate count of one PRF evaluation.
Let $G_{\wedge}^{\mathsf{CapEnc}}\approx 2G_{\wedge}^{\mathsf{PRF}}$ be the AND-gate count for capability encryption
(mask + MAC) when modeled as two PRF calls.
Let $G_{\wedge}^{\mathsf{MAC}}$ be the AND-gate count for one token MAC. This is typically one PRF call, so
$G_{\wedge}^{\mathsf{MAC}}\approx G_{\wedge}^{\mathsf{PRF}}$ up to small formatting overhead.
Let $G_{\wedge}^{\mathsf{sel}}$ and $G_{\wedge}^{\mathsf{fmt}}$ capture small fixed selection and packing/formatting overheads. We summarize our analysis in Table~\ref {tab:glox-complexity-summary}, and the analysis is in Appendix~\ref{app_complexityanalysis}.

\begin{table*}[t]
\centering
\setlength{\tabcolsep}{4pt}
\begin{tabular*}{\textwidth}{@{\extracolsep{\fill}} l l l l @{}}
\toprule
\textbf{Component} & \textbf{GC AND} & \textbf{State (DPF/FSS)} & \textbf{Online comm.} \\
\midrule
\textsf{GetBridge} &
\(G_{\wedge}^{\textsf{GB}}\approx 6G_{\wedge}^{\mathsf{PRF}}+G_{\wedge}^{\mathsf{sel}}+G_{\wedge}^{\mathsf{fmt}}\) &
\(C_{\mathsf{dpfrw}}(M,B),\ T_{\mathsf{dpfrw}}(M,B)\) &
\(2\lambda\,G_{\wedge}^{\textsf{GB}}+C_{\mathrm{OT}}(n_E)\) (bits) \(+\ C_{\mathsf{dpfrw}}(M,B)\) \\
\midrule
\textsf{RedeemDirPIR} &
-- &
-- &
\(2|k_{\mathsf{DPF}}(N_\tau)|+2D+2|\Omega^{\mathsf{pir}}|\) (bytes) \\
\midrule
\textsf{ReportBlocked} (best) &
\(G_{\wedge}^{\textsf{RB,best}}\approx 4G_{\wedge}^{\mathsf{PRF}}+G_{\wedge}^{\mathsf{dedup}}+G_{\wedge}^{\mathsf{fmt}}\) &
\(C_{\mathsf{dpfrw}}(M,B),\ T_{\mathsf{dpfrw}}(M,B)\) &
\(2\lambda\,G_{\wedge}^{\textsf{RB,best}}+C_{\mathrm{OT}}(n_E)\) (bits) \(+\ C_{\mathsf{dpfrw}}(M,B)\) \\
\midrule
\textsf{ReportBlocked} (worst) &
\(G_{\wedge}^{\textsf{RB,worst}}\approx 9G_{\wedge}^{\mathsf{PRF}}+G_{\wedge}^{\mathsf{dedup}}+G_{\wedge}^{\mathsf{sel}}+G_{\wedge}^{\mathsf{fmt}}\) &
\(C_{\mathsf{dpfrw}}(M,B),\ T_{\mathsf{dpfrw}}(M,B)\) &
\(2\lambda\,G_{\wedge}^{\textsf{RB,worst}}+C_{\mathrm{OT}}(n_E)\) (bits) \(+\ C_{\mathsf{dpfrw}}(M,B)\)  \\
& & & \(+\) DirPIR \\
\midrule
\textsf{GroupSplit} &
\(G_{\wedge}^{\textsf{GS}}\approx G_{\wedge}^{\mathsf{cmp}}+G_{\wedge}^{\mathsf{mux}}+G_{\wedge}^{\mathsf{fmt}}\) &
(no extra) &
\(2\lambda\,G_{\wedge}^{\textsf{GS}}+C_{\mathrm{OT}}(n_E)\) (bits) \\
\bottomrule
\end{tabular*}
\caption{Asymptotic costs (Yao GC: Free-XOR, Half-Gates; security \(\lambda\)).}
\label{tab:glox-complexity-summary}
\end{table*}

\subsection{Security Discussion}
\label{subsec:glox-security-discussion}

G--Lox keeps the public interface close to Lox by using anonymous credentials and one-show identifiers, while moving all group-keyed state and bridge-selection logic behind a two-server privacy wall. Under non-collusion of $S_0,S_1$, this lets the system adapt statefully per hidden group without revealing group membership or bridge assignments to the public distributor $D$. Below we summarize how G--Lox defends against common attacks.

\paragraph{Sybil enumeration.}
An adversary can create many identities to repeatedly run \textsf{GetBridge}, collect bridge descriptors, and burn them by
disclosure for blocking. G--Lox inherits Lox-style controls at the public interface: $D$ enforces one-show non-reuse and
rate limits, and invitation-based admission and trust evolution make large-scale Sybils costly in invitation material.
G--Lox further reduces the per-invitation yield. Since the hidden group label $\gamma$ is derived from the invitation
token, identities created from the same invitation material share $\gamma$ and therefore receive the same group-coupled
assignment within an epoch. Thus repeated queries within that invitation-derived group mostly return duplicates, and
enumeration pressure scales primarily with the number of distinct invitation-derived groups the adversary can obtain.

\paragraph{Insider disclosure and bridge burning.}
A malicious but legitimately enrolled user can disclose any bridge they obtain, enabling the adversary to block it. This
cannot be prevented once a client learns a usable descriptor, so the goal is to limit how quickly an insider can obtain
replacements and how much damage one insider can cause. As in rBridge, admission material limits how many independent
identities an insider can cheaply obtain. As in Lox, clients do not receive fresh bridges on demand: rotation is tied to
group evidence of blocking. Concretely, the privacy wall issues a new assignment only when the per-group counter reaches
$\theta$, and only the threshold-crossing report sets $\mathsf{contrib}=1$ for reputation credit. Deduplication prevents
repeated reports from accelerating rotation, so rapid bridge burning requires many distinct insiders or sustained
blocking that also impacts honest users.

\paragraph{Distributor-side linkage.}
The distributor is a high-value vantage point for inferring assignments or social structure. In G--Lox, $D$ never learns
$\gamma$, $\mathsf{tag}$, or the assignment $(\tau,i)$: all assignment material is returned only as end-to-end ciphertexts
to the client, and state evolution is keyed by a hidden tag handled only inside the two-server privacy wall.

\paragraph{Directory scraping and bulk lookup.}
Even with index-private retrieval, an adversary may attempt many lookups to harvest descriptors. G--Lox gates directory
access with a short-lived, one-time authorization token $\Omega^{\mathsf{pir}}$ issued by the privacy wall. Each
directory server enforces token freshness and one-time use, which limits replay and makes bulk scraping require
obtaining many valid authorizations through \textsf{GetBridge}.

\paragraph{Forged reports, replay, and counter inflation.}
An adversary may try to trigger migration by submitting fabricated reports, replaying a stale assignment after rotation,
or repeatedly reporting the same assignment to inflate the threshold counter. G--Lox binds each report to a previously
issued assignment using the opaque token $\Omega^{\mathsf B}$. Inside 2PC, the servers open $\Omega^{\mathsf B}$ under the
secret-shared token key and reject unless it decrypts to a valid tuple $(\tau,e,i)$, preventing forgery or tampering.
They also enforce epoch consistency by accepting only if $e$ matches the current epoch stored in the group record.
Finally, the servers derive a per-user nullifier
$\mathsf{nf}=F_{\mathsf{dedup}}(k_{\mathsf{dedup}},u\|\tau\|e\|i)$ and increment the counter only when $\mathsf{nf}$ is
fresh, so repeated reports for the same assignment by the same user do not accumulate.

\paragraph{Zig-zag discovery pressure.}
Dingledine~\cite{Dingledine11} describes a feedback loop where clients can try many proxies: the adversary blocks one proxy, observes which
clients reconnect, and learns additional proxies those clients try next, gradually expanding coverage. G--Lox reduces the
protocol-level surface for this loop because a client cannot freely choose among many bridges. Each \textsf{GetBridge}
yields a single assignment determined by hidden group state, and within an epoch the assignment is stable for that
hidden group, so repeated requests do not produce a sequence of new bridges that can be learned by successive blocking.
After migration, the next assignment is again produced by the privacy wall and revealed only end-to-end to the client.
This does not eliminate network-layer fingerprinting or traffic-analysis leakage; rather, it removes the mechanism that
would let a client rapidly walk through many bridges in response to blocking. When the adversary learns and blocks a
transport-specific assignment, G--Lox can adapt by switching to a different transport type via the per-group
\textsf{badTypes} state.

\paragraph{Remark for compromised server.}
In the previous systems, all servers are assumed to be semi-honest. In our setting, if one of the servers is compromised, it can then see the full dictionary of bridges. To avoid this, we can enhance the PIR pre-processing that allows each server to hold only partial information of a bridge and reconstruct it in the two-party PIR with small (essentially the same) overhead. Therefore, even if a single server is fully compromised, it still learns neither user information nor the bridge list. 

\begin{figure*}
    \centering
    \includegraphics[trim={0 0.4cm 0 0}, clip, width=1\linewidth]{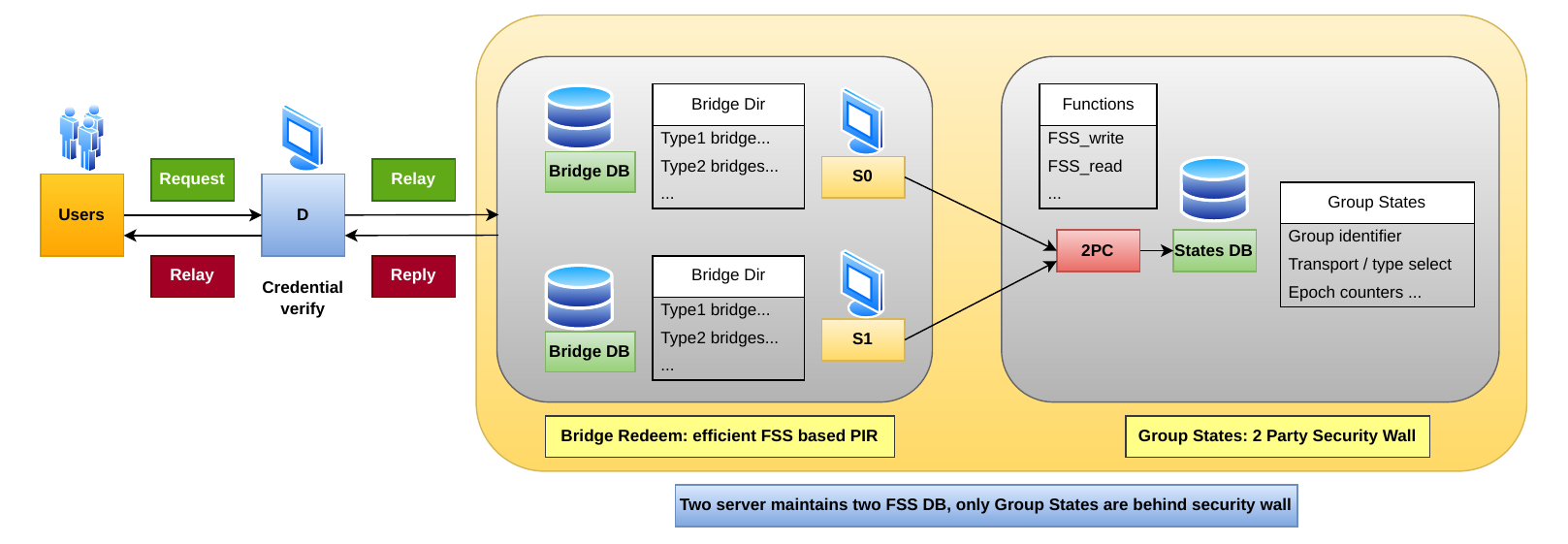}
    \caption{Double FSS-based G-lox workflow}
    \label{fig:fsswf}
\end{figure*}

\section{Evaluation}
\label{sec:eval}

We measure the concrete cost of our two-server DPF-based backend from the running code over real TCP sockets and process memory counters under Linux/WSL. Additionally, we simulate the G-Lox assignment policies and compare them with Lox, Salmon, and rBridge. \footnote{Repository is available as~\url{https://anonymous.4open.science/r/G-lox-5FC5/}.}

\subsection{Evaluation: Micro Benchmark}
\label{subsec:eval-results}
This subsection evaluates the privacy-preserving back-end primitives of G-Lox using a C++17 microbenchmark built with EMP.
Our goal is not to outperform the upstream Lox deployment stack, but to isolate the overhead of private state access and private state-layer policy checks in the two-server setting.

\subsubsection{Setup}
\paragraph{Libraries.}
We implement the microbenchmark in C++17 using the EMP toolkit.
\texttt{emp-tool} provides the AES-based PRG and circuit datatypes, while \texttt{emp-sh2pc} provides a semi-honest Yao garbled-circuit runtime.
We build EMP via CMake \texttt{FetchContent} with pinned versions \texttt{emp-tool v0.2.4}, \texttt{emp-ot v0.2.2}, and \texttt{emp-sh2pc v0.2.2}, compiled in \texttt{Release} mode with AES-NI enabled.

\paragraph{Implementations.}
We implement a real two-server DPF-PIR path for both the state map and the directory.
The state map is replicated across $(S_0,S_1)$, and the client sends one DPF key share to each server.
Each server evaluates its DPF share over the replicated table and returns its response share; XORing the two responses reconstructs the selected record at the client.
For stateful operations, we feed the two server-side DPF response shares into a real EMP semi-honest Yao 2PC between $S_0$ and $S_1$.
This 2PC executes the structured secret-state logic. 

Concretely, the benchmark implements three operation classes.
\textsf{GetBridge} performs blocked/spare handling and a conditional promotion rule on the selected state record.
\textsf{ReportBlocked} updates the secret blockage counter and executes either a migration branch (\texttt{rb\_best}) or a no-spare branch (\texttt{rb\_worst}).
\textsf{Redeem} is implemented as a token-gated state check followed by a directory PIR: the client sends a private token payload, $S_0$ and $S_1$ verify it in 2PC against the selected state record, and the directory query is measured separately.
Directory servers do not run 2PC.

\paragraph{Parameters and Measurement.}
Unless stated otherwise, we fix $\lambda=128$, state-record size $B=128$ bytes, directory descriptor size $D=256$ bytes, and directory size $N=65{,}536$.
We sweep the logical state size $M$ from $2^{10}$ to $2^{16}$.
For each run, we report: (i) client wall-clock runtime; (ii) client-visible TCP traffic; and (iii)inter-server EMP traffic and time from the server summaries. All communication uses loopback TCP under WSL/Linux.

\begin{table}[t]
\centering
\setlength{\tabcolsep}{4pt}
\begin{tabular*}{\columnwidth}{@{\extracolsep{\fill}} r r r r r @{}}
\toprule
$M$ & $S_0$ bytes & $S_1$ bytes & $S_0$ ms & $S_1$ ms \\
\midrule
$1{,}024$  & 417340 & 8010300 & $554.8\pm5.7$  & $427.1\pm3.3$ \\
$2{,}048$  & 417340 & 8010300 & $563.6\pm17.6$ & $432.8\pm7.1$ \\
$4{,}096$  & 417340 & 8010300 & $552.7\pm7.6$  & $429.2\pm6.4$ \\
$8{,}192$  & 417340 & 8010300 & $552.9\pm8.8$  & $429.2\pm5.4$ \\
$16{,}384$ & 417340 & 8010300 & $561.3\pm6.3$  & $437.9\pm5.9$ \\
$32{,}768$ & 417340 & 8010300 & $569.9\pm8.4$  & $442.2\pm6.8$ \\
$65{,}536$ & 417340 & 8010300 & $562.3\pm4.5$  & $436.6\pm5.9$ \\
\bottomrule
\end{tabular*}
\caption{Inter-server EMP cost for G-Lox (5 runs per $M$). Bytes are deterministic; GC runtime is reported as mean $\pm$ std.}
\label{tab:glox-micro-gc}
\end{table}

\begin{table*}[t]
\centering
\setlength{\tabcolsep}{4pt}
\begin{tabular*}{\textwidth}{@{\extracolsep{\fill}} r r r r r r r @{}}
\toprule
$M$ & Client RSS & Client HWM & State max RSS & State max HWM & Dir max RSS & Dir max HWM \\
\midrule
$1{,}024$  & $4.975\pm0.056$ & $4.975\pm0.056$ & $9.793\pm0.000$  & $9.793\pm0.000$  & $20.950\pm0.068$ & $20.950\pm0.068$ \\
$2{,}048$  & $5.025\pm0.056$ & $5.025\pm0.056$ & $9.860\pm0.080$  & $9.860\pm0.080$  & $20.950\pm0.068$ & $20.950\pm0.068$ \\
$4{,}096$  & $5.000\pm0.000$ & $5.000\pm0.000$ & $10.168\pm0.000$ & $10.168\pm0.000$ & $20.900\pm0.056$ & $20.900\pm0.056$ \\
$8{,}192$  & $5.000\pm0.000$ & $5.000\pm0.000$ & $10.666\pm0.003$ & $10.666\pm0.003$ & $20.925\pm0.068$ & $20.925\pm0.068$ \\
$16{,}384$ & $5.000\pm0.000$ & $5.000\pm0.000$ & $11.667\pm0.002$ & $11.667\pm0.002$ & $20.900\pm0.056$ & $20.900\pm0.056$ \\
$32{,}768$ & $5.000\pm0.000$ & $5.000\pm0.000$ & $13.643\pm0.056$ & $13.643\pm0.056$ & $20.975\pm0.056$ & $20.975\pm0.056$ \\
$65{,}536$ & $5.000\pm0.000$ & $5.000\pm0.000$ & $17.631\pm0.082$ & $17.634\pm0.075$ & $20.925\pm0.068$ & $20.925\pm0.068$ \\
\bottomrule
\end{tabular*}
\caption{Peak memory (MB) for the G-Lox microbenchmark (5 runs per $M$), reported as mean $\pm$ std. State and directory columns report the maximum across their two servers for each run.}
\label{tab:glox-micro-memory}
\end{table*}

\begin{table*}[t]
\centering
\setlength{\tabcolsep}{4pt}
\begin{tabular*}{\textwidth}{@{\extracolsep{\fill}} r r r r r r r r r @{}}
\toprule
& \multicolumn{4}{c}{G-Lox} & \multicolumn{4}{c}{Lox baseline} \\
\cmidrule(lr){2-5}\cmidrule(lr){6-9}
$M$ & Sent & Recv & Total & Mean ms
    & Sent & Recv & Total & Mean ms \\
\midrule
$1{,}024$   & 1{,}968 & 1{,}280 & 3{,}248 & $264.9\pm1.1$ & 3{,}876 & 2{,}504 & 6{,}380 & $20.23\pm0.20$ \\
$2{,}048$   & 2{,}076 & 1{,}280 & 3{,}356 & $268.6\pm1.3$ & 3{,}876 & 2{,}504 & 6{,}380 & $19.97\pm0.26$ \\
$4{,}096$   & 2{,}184 & 1{,}280 & 3{,}464 & $271.5\pm2.5$ & 3{,}876 & 2{,}504 & 6{,}380 & $19.99\pm0.23$ \\
$8{,}192$   & 2{,}292 & 1{,}280 & 3{,}572 & $283.2\pm2.9$ & 3{,}876 & 2{,}504 & 6{,}380 & $20.37\pm0.25$ \\
$16{,}384$  & 2{,}400 & 1{,}280 & 3{,}680 & $301.2\pm4.6$ & 3{,}876 & 2{,}504 & 6{,}380 & $20.37\pm0.40$ \\
$32{,}768$  & 2{,}508 & 1{,}280 & 3{,}788 & $240.1\pm3.1$ & 3{,}876 & 2{,}504 & 6{,}380 & $20.22\pm0.30$ \\
$65{,}536$  & 2{,}616 & 1{,}280 & 3{,}896 & $346.3\pm1.7$ & 3{,}876 & 2{,}504 & 6{,}380 & $20.13\pm0.26$ \\
\bottomrule
\end{tabular*}
\caption{Direct comparison of user-visible per-iteration cost for G-Lox and open-source Lox baseline over \texttt{iters}=10. Byte fields are deterministic; runtime is reported as mean $\pm$ std over 5 runs.}
\label{tab:glox-vs-lox-user}
\end{table*}

\subsubsection{Benchmark Report}
We report the benchmark results in Table~\ref{tab:glox-vs-lox-user} and Table~\ref{tab:glox-micro-gc}. The resulting benchmark should be interpreted as a back-end privacy cost study.
It combines real DPF-PIR with real 2PC on structured secret state, but it is still a microbenchmark rather than a full reimplementation of upstream Lox.
Thus, the appropriate baseline question is not whether G-Lox is faster than Lox overall, but what additional latency and communication are incurred by enforcing bridge-distribution policy without revealing the accessed state record to either server alone.
G-Lox is slower than the open-source Lox baseline, but this is not surprising: Lox is not implemented as a two-server DPF/2PC system. The purpose of G-Lox is stronger privacy for state access and state-dependent policy evaluation, not raw performance. The evaluation, therefore, quantifies the privacy overhead of this stronger threat model.

\subsubsection{Baseline Comparison: Lox}
\label{subsec:eval-baseline-lox}
We additionally measured the open-source Lox implementation as a protocol-level baseline using its Rust codebase.\footnote{Repository cloned from \texttt{https://git-crysp.uwaterloo.ca/iang/lox}.}
Because upstream Lox is not implemented as a two-server DPF/PIR system, this comparison is not a primitive-for-primitive match to our EMP-based microbenchmark; instead, it is the closest operation-level baseline for the user-visible workflow.
We instrumented the Lox code to run three operations per iteration:
\textsf{GetBridge}-like (\texttt{open\_invite}),
\textsf{Redeem}-like (\texttt{redeem\_invite}),
and \textsf{ReportBlocked}-like (\texttt{check\_blockage} + \texttt{blockage\_migration}).

For consistency with our sweep, we ran $\texttt{iters}=10$ iterations and varied an equivalent logical state size $M_{\mathrm{eq}} \in \{2^{10},\dots,2^{16}\}$.
Concretely, for each $M_{\mathrm{eq}}$ we instantiated $M_{\mathrm{eq}}/2$ open-invitation buckets and $M_{\mathrm{eq}}/2$ hot-spare buckets in Lox, matching a total of $M_{\mathrm{eq}}$ buckets.
All runs used Ubuntu under WSL2 in \texttt{release} mode. We report serialized request bytes, serialized response bytes, and mean end-to-end per-iteration time measured by the Rust harness.

Across the sweep, the measured serialized traffic is constant at $3{,}876$\,B sent and $2{,}504$\,B received per iteration, for a total of $6{,}380$\,B/iter, while latency remains approximately $19$--$23$\,ms/iter.
This is expected because the Lox protocol messages are fixed-size credential objects and do not scale with a DPF key length.
At the largest point ($M_{\mathrm{eq}}=65{,}536$), the per-operation breakdown is:
\textsf{GetBridge}-like $=1{,}072$\,B and $2.276$\,ms,
\textsf{Redeem}-like $=2{,}256$\,B and $7.337$\,ms,
and \textsf{ReportBlocked}-like $=3{,}052$\,B and $11.825$\,ms. We show the result in Table~\ref{tab:glox-vs-lox-user}.

\subsection{Evaluation: blocking-evasion}
\label{subsec:eval_policy_sim}
To evaluate how G--Lox improves Internet-blocking evasion, we implement and compare the distribution policies 
of G--Lox, rBridge, Lox and Salmon in a policy-level simulator.  The simulator models the distribution decisions
of each of these systems, as well as specific attacks including group-specific blockage, Sybil-driven enumeration, 
and report-based adaptation. 

\paragraph{Baseline grounding.}
Our baseline abstractions are anchored to primary sources and public implementations. For Lox, we model 3-bridge bucket assignment together with reachability-triggered migration. For rBridge, we model credit-priced replacement using \(\phi^- = 45\), matching~\cite{wang2013rbridge}. For Salmon, we model a group-level trust/suspicion policy abstraction guided by the public Salmon codebase, with constants and update rules chosen to capture promotion/demotion based on observed outcomes together with probabilistic issuance denial for low-trust groups.

\paragraph{Simulation model.}
We simulate \(G=128\) hidden groups for 30 days over transports \(\{\textsf{obfs4}, \textsf{snowflake}, \textsf{meek}\}\), averaging results over 14 random seeds. Each group starts with 18--24 users. Daily user arrivals are Poisson, and per-group legitimate daily demand is
\(\max(1,\mathrm{Poisson}(0.26\cdot U_g))\). Each transport begins with 320 bridges. Daily bridge arrivals are Poisson with transport multipliers \((1.10,1.00,0.85)\), and group-specific blocking intensity is sampled from \([0.004,0.018]\) and scaled by transport multipliers \((1.00,1.25,0.85)\). We consider four adversarial strategies: learn and burn, zig-zag, conservative, and blanket-transport, where the last explicitly models transport-wide blocking against selected groups.

\paragraph{Policies.}
\textbf{G-Lox} uses hidden group-adaptive assignment with two deterministic slots per \((g,\tau)\), weighted tuple-level blockage reports, Sybil report weight \(0.15\), decay \(0.68\)/day, trigger threshold \(4.8\), and a requirement that at least one legitimate report be present before migration. \textbf{Lox} is non-group-adaptive: it uses 3-slot bucket-style assignment with global report-triggered migration, threshold \(12.0\), and decay \(0.78\). \textbf{rBridge-like} augments issuance with admission throttling and credit-priced replacement, with price \(45\), gain \(3\) per success, and replacement allowed only when sufficient credit has accumulated. \textbf{Salmon-like} uses group-level trust/suspicion gating. All groups begin from a common initial state; trust is updated from observed success/failure outcomes, with promotion after sustained safe behavior and demotion under persistently poor outcomes; issuance may then be probabilistically denied for sufficiently low-trust groups.

\paragraph{Metrics.}
We report the day 30 bridge-issued rate, day 30 connection success rate, unique bridges exposed to Sybils, and migrations per day. Success is measured over all legitimate requests on day 30 rather than conditioned on issued requests, so selective denial reduces success unless it improves overall bridge availability. Reporting issuance remains important because trust- or reputation-based systems can still trade broader service for lower exposure. We therefore interpret success jointly with issuance. Because migration semantics differ somewhat across systems, we treat migrations/day as a comparative stress indicator rather than a perfectly uniform operational count.

\paragraph{Results.}
In the main stress setting (\texttt{zig\_zag}, $f=0.20$, user-arrival factor $1.6$), G-Lox reaches $76.3\pm3.8\%$ day-30 success with $100.0\%$ issuance. Under the same setting, Lox reaches $38.6\pm3.1\%$ success with $100.0\%$ issuance, rBridge-like reaches $44.3\pm2.4\%$ success with $98.2\%$ issuance, and Salmon-like reaches $58.8\pm3.3\%$ success with $73.9\%$ issuance (100 seeds each). Table~\ref{tab:policy-breakdown-fixed} reports day-30 success as mean $\pm$ standard deviation for fixed-scenario sweeps under the higher-load user-arrival regime (factor $1.8$). In the adversary-type sweep, we fix $f=0.20$ and vary the adversary. In the Sybil-fraction sweep, we fix the adversary to \texttt{zig\_zag} and vary $f$. Because each cell is computed from 100 seeds of a single configuration, the reported standard deviations capture within-scenario seed variability only, avoiding the inflation caused by pooling heterogeneous settings in the earlier aggregated table.

\begin{table}[t]
\centering
\setlength{\tabcolsep}{4pt}
\renewcommand{\arraystretch}{1.08}
\caption{Day-30 success (\%, mean $\pm$ std) from 100 seeds per fixed scenario. Each cell reports variation across seeds only.}
\label{tab:policy-breakdown-fixed}
\begin{tabular*}{\columnwidth}{@{\extracolsep{\fill}} l c c c c @{}}
\toprule
\multicolumn{5}{c}{\textbf{Adversary-type sweep (fixed Sybil fraction $f=0.20$)}} \\
\midrule
System & learn\_burn & zig\_zag & conservative & blanket\_trans \\
\midrule
G-Lox & 38.1$\pm$3.7 & 76.3$\pm$3.8 & 62.0$\pm$4.6 & 70.2$\pm$3.6 \\
Lox & 22.9$\pm$2.8 & 38.6$\pm$3.1 & 42.0$\pm$2.8 & 52.9$\pm$2.7 \\
rBridge-like & 27.9$\pm$2.7 & 44.3$\pm$2.4 & 57.0$\pm$2.6 & 56.5$\pm$3.7 \\
Salmon-like & 53.2$\pm$3.5 & 58.8$\pm$3.3 & 68.2$\pm$3.1 & 51.6$\pm$3.9 \\
\midrule
\multicolumn{5}{c}{\textbf{Sybil-fraction sweep (fixed adversary \texttt{zig\_zag})}} \\
\midrule
System & 0.10 & 0.20 & 0.30 & 0.40 \\
\midrule
G-Lox & 82.5$\pm$3.4 & 76.3$\pm$3.8 & 71.8$\pm$3.5 & 67.0$\pm$3.8 \\
Lox & 47.7$\pm$2.9 & 38.6$\pm$3.1 & 33.7$\pm$3.5 & 31.1$\pm$3.0 \\
rBridge-like & 59.4$\pm$2.6 & 44.3$\pm$2.4 & 27.5$\pm$2.8 & 12.5$\pm$2.0 \\
Salmon-like & 66.7$\pm$2.9 & 58.8$\pm$3.3 & 53.3$\pm$3.0 & 46.9$\pm$3.0 \\
\bottomrule
\end{tabular*}
\end{table}

\paragraph{Takeaway.}
Although G-Lox is more expensive than Lox because it enforces a stronger two-server privacy model, it delivers the strongest robustness among high-issuance systems under group-specific blocking, whereas Salmon-like improves outcomes in part by selectively withholding issuance rather than by preserving broad service.

\section{Conclusion and Future Work}
G-Lox aims to make bridge distribution adaptive to real censorship signals while keeping the public-facing distributor oblivious to users’ bridge assignments and hidden group identifiers. Our core construction is a two-server privacy wall: the two state servers run 2PC to derive the hidden group tag, read/update group-indexed state, and generate opaque bridge tokens, while directory resolution is performed via a lightweight two-server FSS/DPF-based PIR so that neither server under non-collusion assumption learns the queried index. This design supports user-driven blockage reporting, transport-aware rotation, and privacy-preserving group splitting, without leaking stable group identifiers or assignment decisions to the distributor. Our microbenchmark over real TCP sockets makes the concrete cost of G-Lox explicit. Across group-state sizes \(M\in\{2^{10},\ldots,2^{16}\}\), client-visible communication stays in the low-KiB range per end-to-end iteration. 
The measurements show that two-server DPF/FSS-based private access keeps client-edge overhead modest, with only mild growth as \(M\) increases. The dominant remaining cost comes from the backend 2PC required for hidden, state-dependent adaptation. Overall, the evaluation supports the main design intuition behind G-Lox: private state access is practical, and the main scaling pressure lies in the adaptive logic rather than in private retrieval itself.

Regarding future work, we see three natural next steps. First, one can strengthen G-Lox from a privacy-preserving back-end into a more fully metadata-hiding protocol. Our current prototype protects the state-access layer, but the client-visible receive pattern still scales with system parameters, which may leak coarse information unless padded or embedded in a more fully anonymous communication substrate. A natural direction is therefore to combine G-Lox with stronger metadata-hiding or anonymous-communication techniques, such as mixnet-based designs or metadata-private messaging systems, while preserving the bridge-distribution functionality~\cite{piotrowska2017loopix,van2015vuvuzela,lazar2018karaoke, fenske2024bytes}. Second, the prototype can be optimized substantially at the systems level: the benchmark suggests that the dominant cost lies in the 2PC adaptation layer rather than in the private reads themselves, so there is clear room to streamline the state logic, simplify the Boolean circuits, and use batching, pipelining, vectorization, and parallelism to improve throughput. Third, it would be valuable to strengthen the security model beyond the current semi-honest prototype, for example, by moving to maliciously secure MPC back ends or adding stronger integrity checks for the privacy wall~\cite{keller2016mascot,keller2018overdrive}. Taken together, these directions would move G-Lox toward a more end-to-end private, efficient, and robust bridge-distribution system.

\bibliographystyle{ACM-Reference-Format}
\bibliography{refs}  
\appendix

\section{Open Science}

To enable evaluation of the paper's core contributions, we provide anonymized research artifacts for double-blind review, including: (i) the G-Lox prototype implementation, (ii) scripts for reproducing the reported benchmarks and simulations, (iii) configuration files and evaluation parameters, and (iv) documentation for building and running the artifact. Repository is available at~\url{https://anonymous.4open.science/r/G-lox-5FC5}

These materials are shared with the program committee through an anonymous artifact package and an anonymized access link included in the submission. This work does not depend on human-subject data, personal data, or proprietary datasets. We do not require access to production infrastructure to evaluate the paper's main claims.

We may omit deployment-sensitive operational details from the shared materials where release would create unnecessary risk for real-world bridge-distribution systems. Such omissions do not affect the ability of reviewers to assess the paper's scientific contributions, methodology, or reported results.

\section{Ethical Considerations}

This paper studies privacy-preserving bridge distribution for anonymous communication systems. Its aim is defensive: to strengthen privacy and access security in adversarial settings.

We recognize the dual-use nature of this research. While advances in bridge distribution can improve protections for legitimate users, they may also inform adversaries about system assumptions or limitations. Accordingly, we describe the design and analysis at a level sufficient for scientific evaluation while avoiding unnecessary operational detail.

This work does not involve human subjects or personal data. We believe the defensive and privacy-preserving benefits of this research outweigh the risks associated with publication.

\section{Generative AI Usage}
\label{app:ai-usage}

Generative AI tools were used during the preparation of this paper. ChatGPT was used for grammar assistance, and Codex was used for code assistance and implementation support. The authors reviewed, validated, and took responsibility for all technical content, experimental results, code, and claims in the final manuscript.

\section{Lox Protocol review.}
\label{app: low_review}
The Lox system involves three parties. The Lox Authority (LA) is a single central authority that maintains a database of Tor bridges partitioned into buckets, maintains migration tables mapping $(\beta_{\mathrm{FROM}},\beta_{\mathrm{TO}})$ for both trust promotion and blockage migration, and acts as the issuer and verifier for all Lox credentials and tokens. A set of users interacts with the LA over anonymized channels after obtaining their first bridge. Finally, an invitation token distributor (e.g., BridgeDB) hands out open-entry invitation tokens $\Omega^{\mathsf{I}}$. 

The following credential and token types are maintained throughout the system. A \textbf{Lox credential} is
\[
  \Psi^{\mathsf{L}} = (\Phi, t, L, \beta, a, d),
\]
where $\Phi$ is a one-show, unlinkable credential identifier, $t$ is the last update time (e.g., in days), $L \in \{0,1,2,3,4\}$ is the trust level, $\beta$ is the bucket identifier for the user's current bucket, $a$ is the number of invitations the user still holds, and $d$ is a counter of bridges in the user's bucket that have been blocked. An Invitation credential is
\[
  \Psi^{\mathsf{I}} = (\Phi^{\mathsf{I}}, t^{\mathsf{I}}, \beta^{\mathsf{I}}, d^{\mathsf{I}}),
\]
which encodes an invitation from an existing Lox user to a new user; intuitively, $(\beta^{\mathsf{I}}, d^{\mathsf{I}})$ mirror the inviter's bucket and blockage count at issue time. A Bucket reachability credential is
\[
  \Psi^{\mathsf{R}} = (t^{\mathsf{R}}, \beta^{\mathsf{R}}),
\]
issued by the LA each day, attesting that bucket $\beta^{\mathsf{R}}$ was reachable on day $t^{\mathsf{R}}$.
Finally, a Migration token is
\[
  \Omega^{\mathsf{M}} = (\Phi^{\mathsf{L}}, \beta_{\mathrm{FROM}}, \beta_{\mathrm{TO}}, \mathsf{function}),
\]
created by the user after decrypting an encrypted migration table provided by the LA, where $\mathsf{function} \in \{\mathsf{trust},\mathsf{blockage}\}$ indicates whether this is a trust-promotion migration or a blockage migration.

\begin{algorithm}[htbp!]
\caption{\textsf{Lox OpenEntryJoin} (join with open-entry invitation token)}
\label{alg:lox-open-entry}
\begin{algorithmic}[1]
\Require User holds fresh open-entry invitation token $\Omega^{\mathsf I}$.
\Require LA holds bucket state and credential-issuing state.
\Ensure User obtains initial credential $\Psi^{\mathsf L}=(\Phi,t,L{=}0,\beta,a{=}0,d{=}0)$.

\Statex \textbf{User $\to$ LA: \textsf{JoinReq}}
\State Sample $m_\Phi\sample\mathcal{M}$; $(d,D)\gets \mathsf{ElGamal.KeyGen}()$
\State $c_\Phi\gets \mathsf{Enc}_{D}(m_\Phi)$
\State $\pi_{\mathsf{user}}\gets \mathsf{ZKProve}(c_\Phi\text{ encrypts }m_\Phi)$
\State Send $\langle \Omega^{\mathsf I},D,c_\Phi,\pi_{\mathsf{user}}\rangle$

\Statex \textbf{LA: verify \& issue}
\State $\textsf{VerifyFresh}(\Omega^{\mathsf I});\ \textsf{MarkSpent}(\Omega^{\mathsf I})$
\State Verify $\pi_{\mathsf{user}}$
\State Sample $j_\Phi\sample\mathcal{J}$; $\beta\gets\textsf{SampleBucket}()$
\State $(t,L,a,d)\gets(\textsf{now},0,0,0)$
\State $\sigma\gets \textsf{IssueMAC}(D,c_\Phi,j_\Phi,t,L,\beta,a,d)$
\State Send \textsf{JoinResp}$\langle \sigma,j_\Phi,\beta,t\rangle$

\Statex \textbf{User: finalize}
\State $m_\sigma\gets \mathsf{Dec}_{d}(\sigma)$; $\Phi\gets m_\Phi+j_\Phi$
\State Output $\Psi^{\mathsf L}\gets (\Phi,t,L{=}0,\beta,a{=}0,d{=}0)$
\end{algorithmic}
\end{algorithm}

\begin{algorithm}[tbp!]
\caption{\textsf{Lox TrustPromotion+Migration} ($L{=}0\to 1$ into trusted bucket)}
\label{alg:lox-trust-promo}
\begin{algorithmic}[1]
\Require User holds $\Psi^{\mathsf L}_{\text{old}}=(\Phi,t,L{=}0,\beta,a{=}0,d{=}0)$.
\Require LA holds migration table $\{(\beta_{\mathrm{FROM}}^i,\beta_{\mathrm{TO}}^i)\}_i$ and migration-key issuance.
\Ensure User obtains $\Psi^{\mathsf L}_{\text{new}}=(\Phi',t'=\delta,L'=1,\beta'=\beta_{\mathrm{TO}},a'=0,d'=0)$.

\Statex \textbf{Phase 1: eligibility \& migration-key delivery}
\Statex \textbf{User $\to$ LA: \textsf{PromoReq}}
\State $\pi_{\textsf{elig}}\gets \mathsf{ZKProve}[\text{reveal }\Phi;\ \text{hide }(t,\beta);\ \text{eligible}]$
\State Send $\langle \Phi,\pi_{\textsf{elig}}\rangle$
\Statex \textbf{LA}
\State $\textsf{CheckUnused}(\Phi);\ \textsf{MarkPromoUsed}(\Phi)$; verify $\pi_{\textsf{elig}}$
\State $\lambda\gets \textsf{IssueMigKeyCred}(\Phi)$
\State $C_{\textsf{row}}\gets \textsf{EncMigRow}(\lambda;\{(\beta_{\mathrm{FROM}}^i,\beta_{\mathrm{TO}}^i)\}_i)$
\State Send \textsf{PromoResp}$\langle \lambda,C_{\textsf{row}}\rangle$
\Statex \textbf{User}
\State $(\beta_{\mathrm{FROM}},\beta_{\mathrm{TO}})\gets \textsf{DecRow}(\lambda,C_{\textsf{row}})$
\State $\Omega^{\mathsf M}\gets (\Phi,\beta_{\mathrm{FROM}}=\beta,\beta_{\mathrm{TO}},\mathsf{type}=\mathsf{trust})$

\Statex \textbf{Phase 2: migrate \& issue new credential}
\Statex \textbf{User $\to$ LA: \textsf{MigrateReq}}
\State Prepare template with $(t'=\delta,L'=1,a'=0,d'=0,\beta'=\beta_{\mathrm{TO}})$ and fresh $\Phi'$ (jointly derived)
\State $\pi_{\textsf{mig}}\gets \mathsf{ZKProve}[\Omega^{\mathsf M}.\Phi=\Phi;\ \Omega^{\mathsf M}.\beta_{\mathrm{FROM}}=\beta;\ \Omega^{\mathsf M}.\beta_{\mathrm{TO}}=\beta']$
\State Send old cred (reveal $\Phi$; hide $(t,\beta)$), $\Omega^{\mathsf M}$ (hide buckets), new template (hide $(\Phi',\beta')$), and $\pi_{\textsf{mig}}$
\Statex \textbf{LA}
\State Verify $\pi_{\textsf{mig}}$; $\sigma_{\textsf{new}}\gets \textsf{IssueMAC}(\Psi^{\mathsf L}_{\text{new}})$
\State Send \textsf{MigrateResp}$\langle \sigma_{\textsf{new}}\rangle$
\Statex \textbf{User}
\State $\Psi^{\mathsf L}_{\text{new}}\gets \textsf{Finalize}(\sigma_{\textsf{new}})$; store it
\end{algorithmic}
\end{algorithm}

\begin{algorithm}[htbp!]
\caption{\textsf{Lox LevelUp} ($L\to L{+}1$ for $1\le L<4$)}
\label{alg:lox-levelup}
\begin{algorithmic}[1]
\Require Arrays \textsf{DAYS}[1..4], \textsf{INVITATIONS}[1..4], \textsf{MAX\_L}[\(\cdot\)].
\Require User holds $\Psi^{\mathsf L}_{\text{old}}=(\Phi,t,L,\beta,a,d)$ with $1\le L<4$ and reachability cred $\Psi^{\mathsf R}=(t^{\mathsf R},\beta^{\mathsf R})$ for today.
\Ensure User obtains $\Psi^{\mathsf L}_{\text{new}}=(\Phi',t'=\delta,L'=L{+}1,\beta'=\beta,a'=\textsf{INVITATIONS}[L{+}1],d'=d)$.

\Statex \textbf{User $\to$ LA: \textsf{LevelUpReq}}
\State Jointly derive fresh $\Phi'$ with LA; set $\beta'\gets \beta$, $d'\gets d$
\State $\pi_{\textsf{lvl}}\gets \mathsf{ZKProve}[
\beta^{\mathsf R}=\beta \wedge t^{\mathsf R}=\delta \wedge
t+\textsf{DAYS}[L{+}1]\le \delta \le t+\textsf{DAYS}[L{+}1]+511 \wedge
d\le \textsf{MAX\_L}[L{+}1]]$
\State Send reveal $(\Phi,L)$; hide $(t,\beta,a,d)$ and $\beta^{\mathsf R}$; include $\pi_{\textsf{lvl}}$

\Statex \textbf{LA}
\State Verify $\pi_{\textsf{lvl}}$
\State $\textsf{CheckUnused}(\Phi);\ \textsf{MarkLevelUsed}(\Phi)$
\State Set $L'\gets L{+}1$, $t'\gets \delta$, $a'\gets \textsf{INVITATIONS}[L{+}1]$
\State Form $\Psi^{\mathsf L}_{\text{new}}=(\Phi',t',L',\beta',a',d')$; $\sigma_{\textsf{new}}\gets \textsf{IssueMAC}(\Psi^{\mathsf L}_{\text{new}})$
\State Send \textsf{LevelUpResp}$\langle \sigma_{\textsf{new}}\rangle$

\Statex \textbf{User}
\State $\Psi^{\mathsf L}_{\text{new}}\gets \textsf{Finalize}(\sigma_{\textsf{new}})$; store it
\end{algorithmic}
\end{algorithm}

\begin{algorithm}[htbp!]
\caption{\textsf{Lox IssueInvitation} (for $L\ge 2$ and $a>0$)}
\label{alg:lox-issue-inv}
\begin{algorithmic}[1]
\Require User holds $\Psi^{\mathsf L}_{\text{old}}=(\Phi,t,L,\beta,a,d)$ with $L\ge 2$ and $a>0$, and reachability cred $\Psi^{\mathsf R}=(t^{\mathsf R},\beta^{\mathsf R})$ for today.
\Ensure User obtains updated $\Psi^{\mathsf L}_{\text{new}}=(\Phi',t,L,\beta,a-1,d)$ and invitation credential $\Psi^{\mathsf I}=(\Phi^{\mathsf I},t^{\mathsf I}=\delta,\beta^{\mathsf I}=\beta,d^{\mathsf I}=d)$.

\Statex \textbf{User $\to$ LA: \textsf{InviteReq}}
\State Jointly form fresh $\Phi'$ and invitation ID $\Phi^{\mathsf I}$ with LA
\State Prepare hidden updates: $\beta'=\beta$, $L'=L$, $t'=t$, $a'=a-1$, $d'=d$; invitation attrs $\beta^{\mathsf I}=\beta$, $d^{\mathsf I}=d$, $t^{\mathsf I}=\delta$
\State $\pi_{\textsf{inv}}\gets \mathsf{ZKProve}[
\beta^{\mathsf R}=\beta \wedge t^{\mathsf R}=\delta \wedge a>0 \wedge
(\beta',L',t',a',d')=(\beta,L,t,a-1,d) \wedge (\beta^{\mathsf I},d^{\mathsf I})=(\beta,d)]$
\State Send reveal $\Phi$; hide $(t,L,\beta,a,d)$ and $\beta^{\mathsf R}$; include $\pi_{\textsf{inv}}$

\Statex \textbf{LA}
\State Verify $\pi_{\textsf{inv}}$
\State $\textsf{CheckUnused}(\Phi);\ \textsf{MarkInviteUsed}(\Phi)$
\State Issue $\Psi^{\mathsf L}_{\text{new}}=(\Phi',t,L,\beta,a-1,d)$ and $\Psi^{\mathsf I}=(\Phi^{\mathsf I},\delta,\beta,d)$
\State $\sigma_{\textsf{new}}\gets \textsf{IssueMAC}(\Psi^{\mathsf L}_{\text{new}})$; $\sigma_{\textsf I}\gets \textsf{IssueMAC}(\Psi^{\mathsf I})$
\State Send \textsf{InviteResp}$\langle \sigma_{\textsf{new}},\sigma_{\textsf I}\rangle$

\Statex \textbf{User}
\State $\Psi^{\mathsf L}_{\text{new}}\gets \textsf{Finalize}(\sigma_{\textsf{new}})$
\State $\Psi^{\mathsf I}\gets \textsf{Finalize}(\sigma_{\textsf I})$; deliver $\Psi^{\mathsf I}$ to friend
\end{algorithmic}
\end{algorithm}

\begin{algorithm}[htbp!]
\caption{\textsf{Lox RedeemInvitation} (join at $L=1$ using $\Psi^{\mathsf I}$)}
\label{alg:lox-redeem}
\begin{algorithmic}[1]
\Require New user holds invitation credential $\Psi^{\mathsf I}=(\Phi^{\mathsf I},t^{\mathsf I},\beta^{\mathsf I},d^{\mathsf I})$.
\Ensure New user obtains $\Psi^{\mathsf L}=(\Phi,t=\delta,L=1,\beta=\beta^{\mathsf I},a=0,d=d^{\mathsf I})$.

\Statex \textbf{User $\to$ LA: \textsf{RedeemReq}}
\State Prepare template: $\beta\gets \beta^{\mathsf I}$, $d\gets d^{\mathsf I}$, $L\gets 1$, $t\gets \delta$, $a\gets 0$; jointly form fresh $\Phi$ with LA
\State $\pi_{\textsf{red}}\gets \mathsf{ZKProve}[\Psi^{\mathsf I}\text{ valid \& unredeemed} \wedge (t^{\mathsf I}+15\ge \delta)]$
\State Send reveal $\Phi^{\mathsf I}$; hide $(t^{\mathsf I},\beta^{\mathsf I},d^{\mathsf I})$; include $\pi_{\textsf{red}}$

\Statex \textbf{LA}
\State Verify $\pi_{\textsf{red}}$
\State $\textsf{CheckUnused}(\Phi^{\mathsf I});\ \textsf{MarkRedeemed}(\Phi^{\mathsf I})$
\State Issue $\Psi^{\mathsf L}=(\Phi,\delta,1,\beta^{\mathsf I},0,d^{\mathsf I})$; $\sigma_{\textsf L}\gets \textsf{IssueMAC}(\Psi^{\mathsf L})$
\State Send \textsf{RedeemResp}$\langle \sigma_{\textsf L}\rangle$

\Statex \textbf{User}
\State $\Psi^{\mathsf L}\gets \textsf{Finalize}(\sigma_{\textsf L})$; store it
\end{algorithmic}
\end{algorithm}

\section{Complexity Analysis for G-Lox}
\label{app_complexityanalysis}
\paragraph{\textsf{GetBridge}.}
Inside 2PC, the servers compute the hidden tag (1 PRF), derive a type-seed (1 PRF), derive the index $i$ (1 PRF),
generate the bridge token $\Omega^{\mathsf B}$ (modeled as AEAD-style capability encryption: 2 PRFs),
and generate a one-time PIR authorization token $\Omega^{\mathsf{pir}}$ (1 MAC $\approx$ 1 PRF).
Thus the dominant symmetric-crypto count is
\[
N^{\textsf{GB}}_{\mathsf{PRF}}
=
1\ (\mathsf{tag})
+1\ (\textsf{type-seed})
+1\ (i)
+2\ (\Omega^{\mathsf B})
+1\ (\Omega^{\mathsf{pir}})
=6,
\]
and the GC AND-gate count is
\[
G_{\wedge}^{\textsf{GB}}
\approx
6\,G_{\wedge}^{\mathsf{PRF}}
+\;G_{\wedge}^{\mathsf{sel}}
+\;G_{\wedge}^{\mathsf{fmt}}.
\]
Accordingly, the dominant online GC communication is
\[
C^{\mathrm{Yao}}_{\mathrm{GC,online}}(\textsf{GetBridge})
\approx
2\lambda\cdot G_{\wedge}^{\textsf{GB}}\ \text{bits}
\;+\; C_{\mathrm{OT}}(n_E).
\]
Group-state access performs one private read+write of a $B$-byte record, so
\[
C^{\textsf{GB}}_{\mathsf{state}} = C_{\mathsf{dpfrw}}(M,B),
\qquad
T^{\textsf{GB}}_{\mathsf{state}} = T_{\mathsf{dpfrw}}(M,B).
\]
The distributor $D$ only relays an end-to-end ciphertext $C$, so $D$'s online work is dominated by KVAC verification and network relay.

\paragraph{\textsf{RedeemDirPIR}.}
Directory redemption is \emph{not} executed inside the garbled circuit.
Given $(\tau,i,\Omega^{\mathsf{pir}})$, the user generates two DPF keys $(k_0,k_1)\gets \mathsf{DPF.Gen}(i)$ and sends
$(\tau,k_b,\Omega^{\mathsf{pir}})$ to each server $S_b$.
Each server verifies the authorization token (one MAC check) and returns an XOR-share of the $D$-byte descriptor.

\emph{Online communication} per directory PIR is
\[
C_{\mathsf{PIR,online}}
\approx
2\,|k_{\mathsf{DPF}}(N_\tau)|
\;+\;
2D
\;+\;
2|\Omega^{\mathsf{pir}}|
\quad\text{bytes},
\]
where $2D$ bytes is the total size of the two XOR-share replies and the query key size
satisfies $|k_{\mathsf{DPF}}(N_\tau)| = O(\lambda\log N_\tau)$ bits. Note that if the token is transmitted once and cached at the server for a short window, the $2|\Omega^{\mathsf{pir}}|$ term can be
reduced accordingly.

\emph{Per-server work} for redemption consists of (i) one MAC verification plus a spent-set lookup/insert, and
(ii) standard DPF-PIR evaluation over the directory partition:

\[
T_{\mathsf{PIR}}(N_\tau,D)
=
O(1)\
\;+\;
\Theta(N_\tau\cdot D)\ .
\]
Since we already deploy DPF/FSS for group-state access, this redemption step reuses the same primitive; in particular,
the only new component is token gating to prevent directory enumeration.

\paragraph{\textsf{ReportBlocked}: best case.}
Inside 2PC, the servers compute the hidden tag (1 PRF), validate the bridge token $\Omega^{\mathsf B}$
(2 PRFs under $\mathsf{CapDec}$), and derive the dedup nullifier (1 PRF). Thus the dominant PRF count is
\[
N^{\textsf{RB}}_{\mathsf{PRF,best}}
=
1\ (\mathsf{tag})
+\ 2\ (\textsf{token check})
+\ 1\ (\mathsf{nf})
=4.
\]
Accordingly, the AND-gate count is
\[
G_{\wedge}^{\textsf{RB,best}}
\ \approx\
4\,G_{\wedge}^{\mathsf{PRF}}
\;+\;
G_{\wedge}^{\mathsf{dedup}}
\;+\;
G_{\wedge}^{\mathsf{fmt}}.
\]
The dominant online GC communication is
\[
C^{\mathrm{Yao}}_{\mathrm{GC,online}}(\textsf{RB,best})
\approx
2\lambda\cdot G_{\wedge}^{\textsf{RB,best}}\ \text{bits}
\;+\;
C_{\mathrm{OT}}(n_E).
\]

\paragraph{\textsf{ReportBlocked}: worst case.}
If $\mathsf{contrib}=1$, 2PC additionally derives a new type seed (1 PRF), a new index $i_{\mathsf{new}}$ (1 PRF),
generates a fresh bridge token $\Omega^{\mathsf B}_{\mathsf{new}}$ (2 PRFs under capability encryption),
and generates a fresh one-time PIR authorization token $\Omega^{\mathsf{pir}}_{\mathsf{new}}$ (1 MAC $\approx$ 1 PRF).
Thus
\[
\begin{aligned}
N^{\textsf{RB}}_{\mathsf{PRF,worst}}
&=
N^{\textsf{RB}}_{\mathsf{PRF,best}}
+ 1 \ (\textsf{type-seed})
+ 1 \ (i_{\mathsf{new}}) \\
&\quad
+ 2 \ (\Omega^{\mathsf B}_{\mathsf{new}})
+ 1 \ (\Omega^{\mathsf{pir}}_{\mathsf{new}})
= 9 .
\end{aligned}
\]

Accordingly,
\[
G_{\wedge}^{\textsf{RB,worst}}
\ \approx\
9\,G_{\wedge}^{\mathsf{PRF}}
\;+\;
G_{\wedge}^{\mathsf{dedup}}
\;+\;
G_{\wedge}^{\mathsf{sel}}
\;+\;
G_{\wedge}^{\mathsf{fmt}}.
\]
The dominant online GC communication is
\[
C^{\mathrm{Yao}}_{\mathrm{GC,online}}(\textsf{RB,worst})
\approx
2\lambda\cdot G_{\wedge}^{\textsf{RB,worst}}\ \text{bits}
\;+\;
C_{\mathrm{OT}}(n_E),
\]
and the state-backend cost remains one FSS/DPF read+write.

When migration triggers, the user additionally redeems the new assignment via the directory PIR protocol
(\S\ref{alg:glox-redeemdirpir}). This PIR step is outside 2PC; its online communication is analyzed as before.

\paragraph{Split trigger inside 2PC.}
The server-side trigger checks two conditions on the record and, if triggered, samples and stores a fresh seed
$s\sample\{0,1\}^{\lambda}$ and sets a split flag. This is dominated by simple comparisons and conditional assignment
on fixed-size fields already resident in the group record. Accordingly, the AND-gate count for the split logic is
\[
G_{\wedge}^{\textsf{GS}}
\ \approx\ 
G_{\wedge}^{\mathsf{cmp}}
\;+\;
G_{\wedge}^{\mathsf{mux}}
\;+\;
G_{\wedge}^{\mathsf{fmt}},
\]
where $G_{\wedge}^{\mathsf{cmp}}$ accounts for comparing $\mathsf{st}.\textsf{size}$ to $G_{\max}$ and checking the
$\textsf{splitting}$ bit, $G_{\wedge}^{\mathsf{mux}}$ accounts for writing either the old state or the updated state via multiplexing, and $G_{\wedge}^{\mathsf{fmt}}$ is small packing overhead. Notably, this
path does \emph{not} require AES/PRF-style computation; the only randomness needed is a $\lambda$-bit seed $s$, which
is obtained from the servers' local randomness and injected as garbler/evaluator inputs.

Thus, the dominant online GC communication for the split logic is
\[
C^{\mathrm{Yao}}_{\mathrm{GC,online}}(\textsf{GroupSplit})
\approx
2\lambda\cdot G_{\wedge}^{\textsf{GS}}\ \text{bits}
\;+\;
C_{\mathrm{OT}}(n_E),
\]
and in practice is negligible compared to the AES-dominated costs of \textsf{GetBridge}/\textsf{ReportBlocked}.

\end{document}